\documentclass[a4paper,12pt]{article}

\pdfoutput=1
\usepackage{amsmath}
\usepackage{amssymb}
\usepackage{amsfonts}
\usepackage{mathrsfs}
\usepackage{bbm}
\usepackage{graphicx,subfigure,booktabs,adjustbox}
\usepackage[numbers,sort&compress]{natbib}
\usepackage{verbatim}

\usepackage{color}
\usepackage{ulem}
\usepackage{setspace}
\usepackage{multirow}
\usepackage{xcolor,colortbl}
\usepackage[utf8]{inputenc}

\usepackage[colorlinks,
            linkcolor=black,
            filecolor=black,
            anchorcolor=black,
            urlcolor=black,
            citecolor=black,
            bookmarks=false,
            ]{hyperref}

\numberwithin{equation}{section}

\newlength{\dinwidth}
\newlength{\dinmargin}
\makeatletter
\newcommand{\thickhline}{%
    \noalign {\ifnum 0=`}\fi \hrule height 1pt
    \futurelet \reserved@a \@xhline
}
\makeatother

\allowdisplaybreaks

\begin{document}

\title{\bf \Large \boldmath{$B^0\to K^{\ast 0}\mu^+\mu^-$} Decay in the Aligned Two-Higgs-Doublet Model}

\author{
Quan-Yi Hu\footnote{qyhu@mails.ccnu.edu.cn},
Xin-Qiang Li\footnote{xqli@mail.ccnu.edu.cn}\,
and
Ya-Dong Yang\footnote{yangyd@mail.ccnu.edu.cn}\\[15pt]
\small Institute of Particle Physics and Key Laboratory of Quark and Lepton Physics~(MOE), \\
\small Central China Normal University, Wuhan, Hubei 430079, China}

\date{}
\maketitle
\vspace{0.2cm}

\begin{abstract}
{\noindent}In the aligned two-Higgs-doublet model, we perform a complete one-loop computation of the short-distance Wilson coefficients $C_{7,9,10}^{(\prime)}$, which are the most relevant ones for $b\to s\ell^+\ell^-$ transitions. It is found that, when the model parameter $\left|\varsigma_u\right|$ is much smaller than $\left|\varsigma_d\right|$, the charged-scalar contributes mainly to chirality-flipped $C_{9,10}^\prime$, with the corresponding effects being proportional to $\left|\varsigma_d\right|^2$. Numerically, the charged-scalar effects fit into two categories: (A) $C_{7,9,10}^\mathrm{H^\pm}$ are sizable, but $C_{9,10}^{\prime\mathrm{H^\pm}}\simeq0$, corresponding to the (large $\left|\varsigma_u\right|$, small $\left|\varsigma_d\right|$) region; (B) $C_7^\mathrm{H^\pm}$ and $C_{9,10}^{\prime\mathrm{H^\pm}}$ are sizable, but $C_{9,10}^\mathrm{H^\pm}\simeq0$, corresponding to the (small $\left|\varsigma_u\right|$, large $\left|\varsigma_d\right|$) region. Taking into account phenomenological constraints from the inclusive radiative decay $B\to X_s\gamma$, as well as the latest model-independent global analysis of $b\to s\ell^+\ell^-$ data, we obtain the much restricted parameter space of the model. We then study the impact of the allowed model parameters on the angular observables $P_2$ and $P_5'$ of $B^0\to K^{\ast 0}\mu^+\mu^-$ decay, and find that $P_5'$ could be increased significantly to be consistent with the experimental data in case B.
\end{abstract}

\newpage

\section{Introduction}
\label{sec:intro}

The rare $B\to K^{\ast}\ell^+\ell^-$ decays, being the flavour-changing neutral-current (FCNC) processes, do not arise at tree level and are highly suppressed at higher orders within the Standard Model (SM), due to the Glashow-Iliopoulos-Maiani (GIM) mechanism~\cite{Glashow:1970gm}. However, new TeV-scale particles in many extensions of the SM could affect the decay amplitude at a similar level as the SM does. These decays play, therefore, a crucial role in testing the SM and probing various NP scenarios beyond it~\cite{Blake:2016olu}. It is particularly interesting to note that, based on these decays, observables with a very limited sensitivity to hadronic uncertainties can be constructed, enhancing therefore the discovery potential for NP~\cite{Beneke:2001at,Beneke:2004dp,Grinstein:2004vb,Altmannshofer:2008dz,Beylich:2011aq,DescotesGenon:2012zf,Descotes-Genon:2013vna,Gratrex:2015hna}.

Experimentally, several interesting deviations from the SM predictions have been observed in these decays. In 2013, the form-factor-independent angular observable $P'_5$~\cite{DescotesGenon:2012zf,Descotes-Genon:2013vna} of $B^0\to K^{\ast 0}\mu^+\mu^-$ decay was measured by the LHCb collaboration~\cite{Aaij:2013qta}, showing a $3.7\sigma$ disagreement with the SM expectation~\cite{Descotes-Genon:2014uoa,Straub:2015ica,Jager:2012uw,Jager:2014rwa}. Recently, the LHCb collaboration has released new measurements of the angular observables in this decay, based on the dataset of $3~\mathrm{fb}^{-1}$ of integrated luminosity, and still found a $3.4\sigma$ deviation for $P'_5$~\cite{Aaij:2015oid}. Moreover, being in agreement with the LHCb measurements, a deviation with a significance of $2.1\sigma$ was also reported by the Belle collaboration~\cite{Abdesselam:2016llu}. Besides the $P'_5$ anomaly, there are some other slight deviations beyond the $2\sigma$ level, such as the observables $P_2$ in $q^2\in [2,4.3]~\mathrm{GeV}^2$ and $P'_4$ in $q^2\in [14.18,16]~\mathrm{GeV}^2$~\cite{Altmannshofer:2014rta,Descotes-Genon:2015uva,Hurth:2016fbr}. These anomalies have triggered lots of theoretical studies both within the SM and in various NP models~\cite{DescotesGenon:2012zf,Descotes-Genon:2013vna,Gratrex:2015hna,Descotes-Genon:2014uoa,Straub:2015ica,Jager:2012uw,Jager:2014rwa,Hurth:2013ssa,Descotes-Genon:2013wba,Altmannshofer:2013foa,Beaujean:2013soa,Horgan:2013pva,
Hurth:2014vma,Altmannshofer:2014rta,Du:2015tda,Ciuchini:2015qxb,Descotes-Genon:2015uva,Hurth:2016fbr,Meinel:2016grj,Khodjamirian:2010vf,Brass:2016efg,Capdevila:2016ivx,Karan:2016wvu,Ahmed:2016jgv,Chiang:2016qov,Celis:2015eqs,
Boucenna:2016wpr,Crivellin:2016ejn,Barbieri:2016las,Mahmoudi:2016mgr,Crivellin:2015mga,Crivellin:2015lwa,Calibbi:2015kma,Arnan:2016cpy}.

As a minimal extension of the SM scalar sector, the two-Higgs-doublet model (2HDM)~\cite{Lee:1973iz} can easily satisfy the electroweak (EW) precision data and, at the same time, lead to a very rich phenomenology~\cite{Branco:2011iw}. The scalar spectrum consists of two charged scalars $H^\pm$ and three neutral ones $h,\,H$, and $A$, one of which is to be identified with the SM-like Higgs boson found at the LHC~\cite{Aad:2012tfa,Chatrchyan:2012xdj}. The direct search for these additional scalar states would be an important task for high-energy colliders in the next few years. It should be noted that, complementary to the direct searches, indirect constraints on the 2HDM could also be obtained from the rare FCNC decays like $B\to K^{\ast}\ell^+\ell^-$, since these scalars can affect these processes through the penguin and box diagrams. These studies are also very helpful to gain further insights into the scalar sector of supersymmetry and other models that contain similar scalar contents~\cite{Haber:1984rc,Kim:1986ax,Trodden:1998qg}.

In a generic 2HDM, the non-diagonal couplings of neutral scalars to fermions lead to tree-level FCNC interactions, which can be avoided by imposing on the Lagrangian an ad-hoc discrete $\mathcal{Z}_2$ symmetry. Depending on the $\mathcal{Z}_2$ charge assignments to the scalars and fermions, this results in four types of 2HDMs (types I, II, X, Y)~\cite{Branco:2011iw,Gunion:1989we} under the hypothesis of natural flavour conservation (NFC)~\cite{Glashow:1976nt}. In the aligned two-Higgs-doublet model (A2HDM)~\cite{Pich:2009sp}, on the other hand, the absence of tree-level FCNCs is automatically guaranteed by assuming the alignment in flavour space of the Yukawa matrices for each type of right-handed fermions. Interestingly, the A2HDM can recover as particular cases all known specific implementations of the 2HDMs based on $\mathcal{Z}_2$ symmetries. The model is also featured by possible new sources of CP violation beyond that of the Cabibbo-Kobayashi-Maskawa (CKM) matrix~\cite{Cabibbo:1963yz,Kobayashi:1973fv}. These features make the A2HDM very attracting both in high-energy collider physics~\cite{Altmannshofer:2012ar,Bai:2012ex,Barger:2013ofa,Lopez-Val:2013yba,Wang:2013sha,Celis:2013rcs,Celis:2013ixa} and in low-energy flavour physics~\cite{Jung:2010ik,Jung:2010ab,Jung:2012vu,Celis:2012dk,Duarte:2013zfa,Jung:2013hka,Li:2014fea,Ilisie:2015tra,Abbas:2015cua,Han:2015yys,Wang:2016rvz}.

In this paper, we will study the decay $B^0\to K^{\ast 0}\mu^+\mu^-$ in the A2HDM. Our paper is organized as follows: In section~\ref{sec:A2HDM}, we give a brief overview of the A2HDM, focusing mainly on the scalar and Yukawa sectors. In section~\ref{sec:calculate}, a complete one-loop computation of the short-distance (SD) Wilson coefficients $C_{7,9,10}^{(\prime)}$ is presented, and the final analytical expressions are given both within the SM and in the A2HDM. The angular observables of $B^0\to K^{\ast 0}\mu^+\mu^-$ decay are also introduced in this section. In section~\ref{sec:results}, taking into account phenomenological constraints from the inclusive radiative decay $B\to X_s\gamma$ and the latest model-independent global analysis of $b\to s\ell^+\ell^-$ data, we study the impact of the allowed model parameters on the angular observables $P_2$ and $P_5'$ of $B^0\to K^{\ast 0}\mu^+\mu^-$ decay. Finally, our conclusions are made in section~\ref{sec:conclusion}. Some relevant functions for the Wilson coefficients are collected in the appendices.

\section{The aligned two-Higgs doublet model}
\label{sec:A2HDM}

We consider the minimal version of 2HDM, which is invariant under the SM gauge group and includes, besides the SM matter and gauge fields, two complex scalar $\mathrm{SU\left(2\right)_{L}}$ doublets,
\begin{equation}
\phi_a^T\left(x\right)=\frac{\mathrm{e}^{i\theta_a}}{\sqrt{2}}\left(\sqrt{2}\varphi^+_a,
\,v_a+\rho_a+i\eta_a\right), \qquad \left(a=1,\,2\right)\,,
\end{equation}
with the hypercharge $Y=1/2$. The neutral components of the two scalar doublets acquire the vacuum expectation values (VEVs) $\langle 0|\phi_a^T\left(x\right)|0\rangle = \left(0, v_a\mathrm{e}^{i\theta_a}/\sqrt{2}\right)$. Through an appropriate $\mathrm{U(1)}_Y$ transformation, one can enforce $\theta_1=0$ and leave the relative phase $\theta=\theta_2-\theta_1$ as physical. Using further a global $\mathrm{SU\left(2\right)}$ transformation in the scalar space, one can rotate the original scalar basis to the so-called Higgs basis~\cite{Davidson:2005cw,Haber:2006ue,Haber:2010bw},
\begin{equation}
\left( \begin{array}{c} \Phi_1 \\ -\Phi_2 \end{array} \right) \equiv
\left( \begin{array}{cc} \cos\beta & \sin\beta \\ \sin\beta & -\cos\beta \end{array} \right)\,
\left( \begin{array}{c} \phi_1 \\  \mathrm{e}^{-i\theta}\phi_2 \end{array} \right)\; ,
\end{equation}
where the rotation angle (clockwise) $\tan\beta=v_2/v_1$. In the new basis, only the scalar doublet $\Phi_1$ gets a nonzero VEV $\langle 0|\Phi_1^T\left(x\right)|0\rangle = \left(0, v/\sqrt{2}\right)$, with $v=\sqrt{v_1^2+v_2^2}=(\sqrt{2} G_F)^{-1/2} \simeq 246~\mathrm{GeV}$, and the two scalar doublets are now parametrized, respectively, by~\cite{Pich:2009sp}
\begin{equation} \label{eq:Higgsbasis}
 \Phi_1=\left( \begin{array}{c} G^+ \\ \frac{1}{\sqrt{2}}\, (v+S_1+iG^0) \end{array} \right) \; ,
 \qquad
 \Phi_2 = \left( \begin{array}{c} H^+ \\ \frac{1}{\sqrt{2}}\, (S_2+iS_3) \end{array} \right) \; ,
\end{equation}
where $G^\pm$ and $G^0$ denote the massless Goldstone fields to be eaten by the $W^\pm$ and $Z^0$ gauge bosons, respectively. The remaining five physical degrees of freedom are given by the two charged fields $H^\pm(x)$ and the three neutral ones $\varphi^0_i(x) =\{h(x), H(x), A(x)\}=\mathcal{R}_{ij}S_j$, where $\mathcal{R}$ is an orthogonal matrix obtained after diagonalizing the mass terms in the scalar potential. Generally, none of these three neutral scalars can have a definite CP quantum number.

\subsection{Scalar sector}
\label{sec:Scalar}

The most general scalar potential for the two doublets $\Phi_1$ and $\Phi_2$ that is allowed by the EW gauge symmetry can be written as~\cite{Davidson:2005cw,Haber:2006ue,Haber:2010bw}:
\begin{align} \label{eq:potential}
V &= \mu_{1}\;\left(\Phi_{1}^{\dagger}\Phi_{1}\right)\,+\,\mu_{2}\;\left(\Phi_{2}^{\dagger}\Phi_{2}\right)\,
+\,\left[\mu_{3}\;\left(\Phi_{1}^{\dagger}\Phi_{2}\right)\,+\,\mu_{3}^{*}\;\left(\Phi_{2}^{\dagger}\Phi_{1}
\right)\right]\nonumber\\[0.2cm]
& +\lambda_{1}\,\left(\Phi_{1}^{\dagger}\Phi_{1}\right)^{2}\,+\,\lambda_{2}\,\left(\Phi_{2}^{\dagger}\Phi_{2}\right)^{2}\,
+\,\lambda_{3}\,\left(\Phi_{1}^{\dagger}\Phi_{1}\right)\left(\Phi_{2}^{\dagger}\Phi_{2}\right)\,
+\,\lambda_{4}\,\left(\Phi_{1}^{\dagger}\Phi_{2}\right)\left(\Phi_{2}^{\dagger}\Phi_{1}\right)\nonumber\\[0.2cm]
& +\left[\left(\lambda_{5}\;\Phi_{1}^{\dagger}\Phi_{2}\,+\,\lambda_{6}\;\Phi_{1}^{\dagger}\Phi_{1}\,+\,\lambda_{7}\;
\Phi_{2}^{\dagger}\Phi_{2}\right)\left(\Phi_{1}^{\dagger}\Phi_{2}\right)\,+\,\mathrm{h.c.}\right]\,.
\end{align}
The hermiticity of the potential requires the parameters $\mu_{1,2}$ and $\lambda_{1,2,3,4}$ to be real, while $\mu_3$ and $\lambda_{5,6,7}$ could be generally complex. The minimization condition imposes the relations $\mu_1 = -\lambda_1 v^2$ and $\mu_3 = -\frac{1}{2}\,\lambda_6\, v^2$. Since only the relative phases among $\lambda_{5,6,7}$ are physical, the scalar potential is finally fully characterized by eleven real parameters, $v$, $\mu_2$, $\lambda_{1,2,3,4}$, $|\lambda_{5,6,7}|$, $\mathrm{arg}(\lambda_5\lambda_6^*)$ and $\mathrm{arg}(\lambda_5\lambda_7^*)$, four of which can be determined by the scalar masses $M_{H^\pm,\,h,\,H,\,A}$. Explicitly, inserting Eq.~(\ref{eq:Higgsbasis}) into Eq.~(\ref{eq:potential}) and imposing the minimization condition, one gets $M_{H^\pm}^2=\mu_2+\frac{1}{2}\lambda_3v^2$, and the mass-squared matrix $\mathcal{M}^{2}$ of $S_{1,2,3}$ fields in terms of $v$ and $\lambda_i$. Using the orthogonal matrix $\mathcal{R}$, one can then obtain the masses of the three neutral scalars, $\mathcal{R}\,\mathcal{M}^2\,\mathcal{R}^T = \mathrm{diag}\left( M_h^2, M_H^2,M_A^2\right)$.

In the CP-conserving limit, $\lambda_{5,6,7}$ are all real and the neutral scalars are CP eigenstates. The CP-odd scalar $A$ corresponds to $S_3$, with the mass given by $M_A^2=M_{H^\pm}^2+ v^2\left(\frac{\lambda_4}{2}-\lambda_5\right)$, while the two CP-even scalars $h$ and $H$ are orthogonal combinations of $S_1$ and $S_2$,
\begin{align}
\left(\begin{array}{c} h\\ H \end{array} \right)\; = \;
\left(\begin{array}{cc} \cos{\tilde\alpha} & \sin{\tilde\alpha} \\ -\sin{\tilde\alpha} & \cos{\tilde\alpha} \end{array}\right)\;
\left(\begin{array}{c} S_1\\ S_2 \end{array}\right) \,,
\end{align}
where the mixing angle $\tilde\alpha$ is determined by
\begin{equation}
\tan{\tilde\alpha}\; =\; \frac{M_h^2 - 2\lambda_1 v^2}{v^2\lambda_6}
\; =\; \frac{v^2\lambda_6}{2\lambda_1 v^2- M_H^2}\,.
\end{equation}
The masses of the two neutral scalars are given, respectively, by $M_h^2=\frac{1}{2}\left(\Sigma-\Delta\right)$ and $M_H^2=\frac{1}{2}\left(\Sigma+\Delta\right)$, where
\begin{align}
& \Sigma \, =\, M_{H^\pm}^2\, +\, v^2\,\left(2\,\lambda_1 +\frac{\lambda_4}{2}+ \lambda_5\right)\,,\nonumber \\[0.2cm]
& \Delta\, =\,\sqrt{\left[M_{H^\pm}^2\, +\, v^2\,\left(-2\,\lambda_1 +\frac{\lambda_4}{2}+ \lambda_5\right) \right]^2 + 4 v^4 \lambda_6^2}
\, = \, - \frac{2 v^2\lambda_6}{\sin{(2\tilde\alpha)}}\, .
\end{align}
Here $M_h \leqslant M_H$ by convention and the SM limit is recovered when $\tilde\alpha=0$.

\subsection{Yukawa sector}
\label{sec:Yukawa}

The Yukawa Lagrangian of the 2HDM is most generally given by~\cite{Pich:2009sp,Branco:2011iw}
\begin{equation} \label{eq:Yukawa}
\mathcal L_Y  =  -\left[
\bar{Q}_L' (\Gamma_1 \phi_1 +\Gamma_2 \phi_2) d_R' + \bar{Q}_L' (\Delta_1 \tilde{\phi}_1 +\Delta_2 \tilde{\phi}_2) u_R'  + \bar{L}_L' (\Pi_1 \phi_1 + \Pi_2 \phi_2) \ell_R'\right]
 + \mathrm{h.c.}  \; ,
\end{equation}
where $\tilde{\phi}_a(x)\equiv i\tau_2\phi_a^{\ast}(x)$ are the charge-conjugated fields with $Y=-\frac{1}{2}$, $\bar{Q}_L'$ and $\bar{L}_L'$ are the left-handed quark and lepton doublets, and $u'_R$, $d'_R$ and $\ell'_R$ the corresponding right-handed singlets, in the weak-interaction basis. All fermionic fields are written as 3-dimensional vectors and the couplings $\Gamma_a$, $\Delta_a$ and $\Pi_a$ are therefore $3\times 3$ complex matrices in flavour space.

Transforming to the Higgs basis, Eq.~(\ref{eq:Yukawa}) becomes
\begin{equation}
 \mathcal{L}_Y = -\frac{\sqrt{2}}{v}\,\Big[\bar{Q}'_L (M'_d \Phi_1 + Y'_d \Phi_2) d'_R + \bar{Q}'_L (M'_u \tilde{\Phi}_1 + Y'_u \tilde{\Phi}_2) u'_R + \bar{L}'_L (M'_\ell \Phi_1 + Y'_\ell \Phi_2) \ell'_R \Big] + \mathrm{h.c.} \,,
\end{equation}
where
\begin{align}
M_d'&=\frac{1}{\sqrt{2}}\left(v_1\Gamma_1+v_2\Gamma_2\mathrm{e}^{i\theta}\right)\,,
&
Y_d'&=\frac{1}{\sqrt{2}}\left(-v_2\Gamma_1+v_1\Gamma_2\mathrm{e}^{i\theta}\right)\,,
\\[0.2cm]
M_u'&=\frac{1}{\sqrt{2}}\left(v_1\Delta_1+v_2\Delta_2\mathrm{e}^{-i\theta}\right)\,,
&
Y_u'&=\frac{1}{\sqrt{2}}\left(-v_2\Delta_1+v_1\Delta_2\mathrm{e}^{-i\theta}\right)\,,
\\[0.2cm]
M_\ell'&=\frac{1}{\sqrt{2}}\left(v_1\Pi_1+v_2\Pi_2\mathrm{e}^{i\theta}\right)\,,
&
Y_\ell'&=\frac{1}{\sqrt{2}}\left(-v_2\Pi_1+v_1\Pi_2\mathrm{e}^{i\theta}\right)\,.
\end{align}
In general, the Yukawa matrices $M_f'$ and $Y_f'$~($f=u,d,\ell$) cannot be simultaneously diagonalized in flavour space. Thus, in the mass-eigenstate basis, with diagonal fermion mass matrices $M_f$, the corresponding Yukawa matrices $Y_f$ remain non-diagonal, giving rise to tree-level FCNC interactions. The unwanted tree-level FCNCs can be eliminated by requiring the alignment in flavour space of the Yukawa matrices~\cite{Pich:2009sp}:
\begin{align}
& \Gamma_2 = \xi_d\, \mathrm{e}^{-i\theta} \,\Gamma_1 \; , \qquad \Delta_2=\xi_u^\ast\, \mathrm{e}^{i\theta}\Delta_1\; ,  \qquad \Pi_2=\xi_\ell\, \mathrm{e}^{-i\theta} \,\Pi_1 \; , \nonumber \\[0.2cm]
& Y_{d,\ell}=\varsigma_{d,\ell}\, M_{d,\ell}\, ,
\qquad
  Y_u=\varsigma^\ast_u\, M_u\, ,
\qquad
 \varsigma_f \equiv \frac{\xi_f-\tan{\beta}}{1+\xi_f\tan{\beta}}\,,
\end{align}
where $\xi_f$~($\varsigma_f$) are arbitrary complex parameters and could introduce new sources of CP violation beyond that of the CKM matrix.

The interactions of the charged scalar with the fermion mass-eigenstate fields then read
\begin{equation}
 \mathcal{L}_{H^\pm}=- \frac{\sqrt{2}}{v}\, H^+\, \bigg\{\bar{u} \Big[\varsigma_d\,V_{\mathrm{CKM}} M_d P_R - \varsigma_u\,M_u^{\dagger} V_{\mathrm{CKM}} P_L\Big] d + \varsigma_\ell\,\bar{\nu} M_\ell P_R \ell \bigg\}+ \mathrm{h.c.} \,,
\end{equation}
where $P_{L(R)}\equiv (1\mp\gamma_5)/2$ is the left~(right)-handed chirality projector, and $V_{\mathrm{CKM}}$ the CKM matrix~\cite{Cabibbo:1963yz,Kobayashi:1973fv}. Here we did not give the neutral scalar sector~\cite{Pich:2009sp} in $\mathcal{L}_Y$ or the FCNC local structures induced beyond tree-level~(quantum corrections)~\cite{Jung:2010ik}, because their effects are highly suppressed by the muon mass in the decay $B^0\to K^{\ast 0}\mu^+\mu^-$. The usual NFC models~\cite{Branco:2011iw,Gunion:1989we}, with discrete $\mathcal{Z}_2$ symmetries, are recovered for particular values of $\varsigma_f$, as shown in Table~\ref{tab:models}.

\begin{table}[t]
\begin{center}
\caption{\small The one-to-one correspondence between different specific choices of the couplings $\varsigma_f$ and the 2HDMs based on discrete $\mathcal{Z}_2$ symmetries.}
\vspace{0.2cm} \tabcolsep 0.2in
\renewcommand\arraystretch{1.3}
\begin{tabular}{|c|c|c|c|}
\hline \rowcolor{lightgray}
Model & $\varsigma_d$ & $\varsigma_u$ & $\varsigma_\ell$  \\
\hline
Type I  & $\cot{\beta}$ &$\cot{\beta}$ & $\cot{\beta}$ \\ \rowcolor{lightgray}
Type II & $-\tan{\beta}$ & $\cot{\beta}$ & $-\tan{\beta}$ \\
Type X  & $\cot{\beta}$ & $\cot{\beta}$ & $-\tan{\beta}$ \\ \rowcolor{lightgray}
Type Y  & $-\tan{\beta}$ & $\cot{\beta}$ & $\cot{\beta}$ \\
Inert  & 0 & 0 & 0 \\
\hline
\end{tabular}
\label{tab:models}
\end{center}
\end{table}

\section{$\boldsymbol{B^0\to K^{\ast 0}\mu^+\mu^-}$ in the A2HDM}
\label{sec:calculate}

\subsection{Effective weak Hamiltonian}

The rare decay $B^0\to K^{\ast 0}\mu^+\mu^-$ proceeds through the loop diagrams both within the SM and in the A2HDM. When the heavy degrees of freedom, including the top quark, the weak gauge bosons, as well as the charged scalars, have been integrated out, we obtain the low-energy effective weak Hamiltonian governing the decay~\cite{Buchalla:1995vs,Altmannshofer:2008dz}:
\begin{eqnarray} \label{eq:Heff}
\mathcal{H}_{\mathrm{eff}}=-\frac{4G_{F}}{\sqrt{2}}V_{tb}V_{ts}^{\ast}\sum_{i}\left(C_{i}O_{i} +C_{i}^{\prime}O_{i}^{\prime}\right)\,,
\end{eqnarray}
where $G_F$ is the Fermi coupling constant. Here we neglect the doubly Cabibbo-suppressed (proportional to $V_{ub}V_{us}^{\ast}$) contributions to Eq.~\eqref{eq:Heff}, and focus only on the operators~\cite{Altmannshofer:2008dz}:
\begin{align}
O_7 &= \frac{e}{16\pi^2}\bar m_b\left(\bar s\sigma^{\mu\nu}P_R b\right)F_{\mu\nu}\,, &
O^{\prime}_7 &= \frac{e}{16\pi^2}\bar m_b\left(\bar s\sigma^{\mu\nu}P_L b\right)F_{\mu\nu}\,,
\\[0.2cm]
O_9 &= \frac{e^2}{16\pi^2}\left(\bar s\gamma^\mu P_L b\right)\left(\bar \mu\gamma_\mu\mu\right)\,, &
O^{\prime}_9 &= \frac{e^2}{16\pi^2}\left(\bar s\gamma^\mu P_R b\right)\left(\bar \mu\gamma_\mu\mu\right)\,,
\\[0.2cm]
O_{10} &= \frac{e^2}{16\pi^2}\left(\bar s\gamma^\mu P_L b\right)\left(\bar\mu\gamma_\mu\gamma_5 \mu\right)\,, &
O^{\prime}_{10} &= \frac{e^2}{16\pi^2}\left(\bar s\gamma^\mu P_R b\right)\left(\bar\mu\gamma_\mu\gamma_5 \mu\right)\,,
\end{align}
where $\bar m_b=\bar m_b(\mu)$ denotes the $b$-quark running mass in the $\mathrm{\overline{MS}}$ scheme.

Within the SM, the electromagnetic dipole operator $O_7$ and the semileptonic operators $O_{9,10}$ play the leading role in the decay $B^0\to K^{\ast 0}\mu^+\mu^-$. Besides modifying the values of the SD Wilson coefficients $C_{7,9,10}$, the charged-scalar contributions could also make the chirality-flipped operators $O_{7,9,10}^{\prime}$ defined above to contribute in a significant manner, especially in some regions of the parameter spaces discussed later.

The SD Wilson coefficients $C_i(\mu)$ and $C_i^{\prime}(\mu)$ can be obtained firstly at the matching scale $\mu_W\sim M_W$ perturbatively, by requiring equality of the one-particle irreducible Green functions calculated in the full and in the effective theory~\cite{Buchalla:1995vs}. Using the renormalization group equation, one can then get $C_i(\mu)$ and $C_i^{\prime}(\mu)$ at the lower scale $\mu_b\sim m_b$. During the calculation, the limit $\bar m_{u,c}\to 0$ and the unitarity of the CKM matrix have been used. For simplicity, we introduce the mass ratios:
\begin{eqnarray}
x_t=\frac{\bar m^2_t(\mu_W)}{M^2_W},\qquad  y_t=\frac{\bar m^2_t(\mu_W)}{M^2_{H^\pm}}\,.
\end{eqnarray}
Details of the computational method could be found, for example, in refs.~\cite{Li:2014fea,Buchalla:1995vs}.

\subsection{Wilson coefficients in the SM}

In the SM, the one-loop penguin and box diagrams have been calculated both in the Feynman~($\xi=1$) and in the unitary~($\xi=\infty$) gauge~\cite{Inami:1980fz,Misiak:1992bc,Deshpande:1981zq,Deshpande:1982mi,Chia:1983hd,Chia:1985dx,Chia:1989gh,Wu:2006sp,He:2009rz}, denoted by the subscript `F' and `U', respectively. The different contributions to $C_i^{\mathrm {SM}}(\mu_W)$ can be split into the following forms:
\begin{eqnarray} \label{eq:C7SM}
C_{7}^{\mathrm {SM}} & = & C_{7}^{\gamma\text{-penguin}},\\ \label{eq:C9SM}
C_{9}^{\mathrm {SM}} & = & C_{9}^{W\text{-box}}+C_{9}^{Z\text{-penguin}}+C_{9}^{\gamma\text{-penguin}},\\ \label{eq:C10SM}
C_{10}^{\mathrm {SM}} & = & C_{10}^{W\text{-box}}+C_{10}^{Z\text{-penguin}}\,,
\end{eqnarray}
where the corresponding parts resulting from the $W$-box, $Z$-penguin and $\gamma$-penguin diagrams are given, respectively, by
\begin{align}
&C_{9,\mathrm{F(U)}}^{W\text{-box}} = -\frac{B_{0,\mathrm{F(U)}}}{\sin^{2}\theta_{W}}\,, &
&C_{10,\mathrm{F(U)}}^{W\text{-box}} = \frac{B_{0,\mathrm{F(U)}}}{\sin^{2}\theta_{W}}\,,
\\[0.1cm]	
&C_{9,\mathrm{F(U)}}^{Z\text{-penguin}} = \left(-4+\frac{1}{\sin^{2}\theta_{W}}\right)C_{0,\mathrm{F(U)}}\,, &
&C_{10,\mathrm{F(U)}}^{Z\text{-penguin}} = -\frac{C_{0,\mathrm{F(U)}}}{\sin^{2}\theta_{W}}\,,
\\[0.1cm]
&C_{7,\mathrm{F(U)}}^{\gamma\text{-penguin}} = -\frac{1}{2}D_{0,\mathrm{F(U)}}'\,, &
&C_{9,\mathrm{F(U)}}^{\gamma\text{-penguin}} = -D_{0,\mathrm{F(U)}}+\frac{4}{9}\,,
\end{align}
where $\theta_W$ is the weak mixing angle, and the Inami-Lim functions~\cite{Inami:1980fz} are defined as
\begin{align}\label{eq:fi_feynman}
B_{0,\mathrm{F}}=F_{1}\left(x_{t}\right),\quad C_{0,\mathrm{F}}=F_{3}\left(x_{t}\right),\quad D_{0,\mathrm{F}}'=F_{6}\left(x_{t}\right),\quad D_{0,\mathrm{F}}=-\frac{4}{9}F_{0}\left(x_{t}\right)+F_{5}\left(x_{t}\right),
\end{align}
in the Feynman gauge, and
\begin{align}\label{eq:fi_unitary}
&B_{0,\mathrm{U}} = -\frac{x_{t}}{16}L_{\epsilon}+F_{4}\left(x_{t}\right)\,, &
&C_{0,\mathrm{U}} = -\frac{x_{t}}{16}L_{\epsilon}-F_{1}\left(x_{t}\right)+F_{3}\left(x_{t}\right)+F_{4}\left(x_{t}\right)\,,\nonumber
\\[0.2cm]	
&D_{0,\mathrm{U}}' = F_{6}\left(x_{t}\right)\,,&
&D_{0,\mathrm{U}} = \frac{x_{t}}{4}L_{\epsilon}-\frac{4}{9}F_{0}\left(x_{t}\right)+4F_{1}\left(x_{t}\right)-4F_{4}\left(x_{t}\right)+F_{5}\left(x_{t}\right)\,,
\end{align}
in the unitary gauge. Here we introduce the notation $L_{\epsilon}\equiv\frac{1}{\epsilon}+\log\left(\frac{\mu_W^{2}}{M_{W}^{2}}\right)$, where $\epsilon=(4-d)/2$ is the dimensional regulator of ultraviolet divergence. Explicit expressions of the basic functions $F_i(x)$ are given by Eqs.~(\ref{eq:BF1})--(\ref{eq:BF9}). While each piece on the right-hand side of Eqs.~(\ref{eq:C9SM}) and (\ref{eq:C10SM}) depends obviously on $\epsilon$ in the unitary gauge, due to the longitudinal components of the $W^\pm$, $Z^0$ and off-shell photon propagators, the physical quantities $C^{\mathrm{SM}}_{7,9,10}$ are indeed free of $\epsilon$ and are independent of the EW gauge fixings. For a recent review of higher-order corrections to $C^{\mathrm{SM}}_{7,9,10}$, the readers are referred to ref.~\cite{Buras:2011we}.

\subsection{Wilson coefficients in the A2HDM}

\begin{figure}[t]
  \centering
  \includegraphics[width=0.9\textwidth]{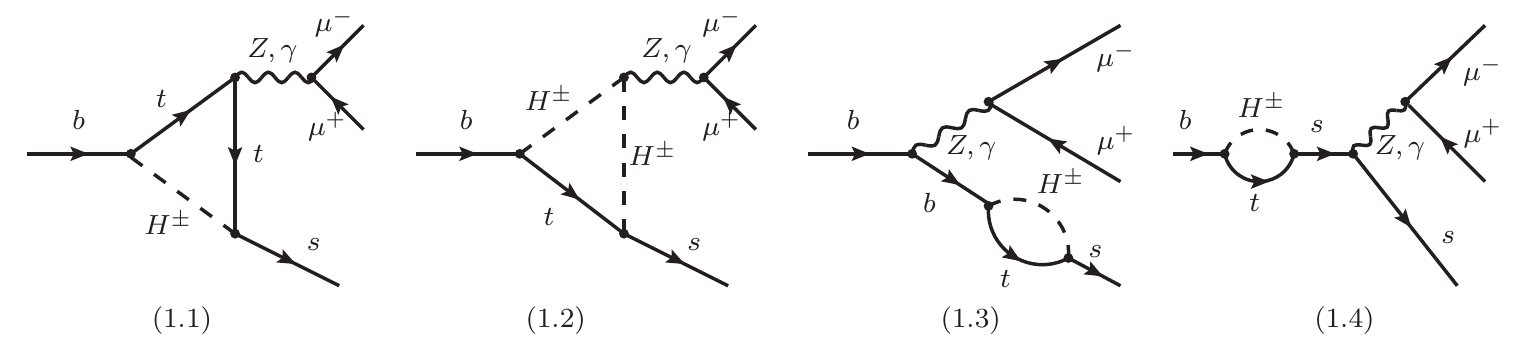}
  \caption{\small $Z$- and $\gamma$-penguin diagrams involving the charged-scalar exchanges in the A2HDM.}
  \label{fig:ZgA2HDM}
\end{figure}

In the A2HDM, the charged-scalar exchanges lead to additional contributions to $C_{7,9,10}$ and could also make the chirality-flipped operators $O^{\prime}_{7,9,10}$ to contribute in a significant manner, through the $Z^0$- and $\gamma$-penguin diagrams shown in Figure~\ref{fig:ZgA2HDM}. Since we have neglected the light lepton mass, there is no contribution from the SM $W$-box diagrams with the $W^{\pm}$ bosons replaced by the charged scalars $H^{\pm}$.

For each Feynman diagram shown in Figure~\ref{fig:ZgA2HDM}, the contributions are identical in the two gauges. The total Wilson coefficients $C_{7,9,10}$ are split into two parts, one is from the SM contributions $C_{7,9,10}^\mathrm{SM}$, and the other from the charged-scalar ones $C_{7,9,10}^\mathrm{H^\pm}$. For the chirality-flipped operators, $C_{7,9,10}^{\prime}=C_{7,9,10}^{\prime\mathrm{H^\pm}}$, because the SM contributions are well suppressed by the factor $\bar{m}_s/\bar{m}_b$. For convenience, we decompose these new contributions in such a way to render explicit their dependence on the couplings $\varsigma_u$ and $\varsigma_d$:
\begin{align}
&C_{7}^\mathrm{H^\pm}=\left|\varsigma_{u}\right|^{2}C_{7,\,\mathrm{uu}} +\varsigma_{d}\varsigma_{u}^{\ast}C_{7,\,\mathrm{ud}}\,, \label{eq:WC7inA2HDM} \\[0.2cm]
&C_{9}^\mathrm{H^\pm}=\left|\varsigma_{u}\right|^{2}C_{9,\,\mathrm{uu}}\,, \label{eq:WC9inA2HDM} \\[0.2cm]
&C_{10}^\mathrm{H^\pm}=\left|\varsigma_{u}\right|^{2}C_{10,\,\mathrm{uu}}\,,\label{eq:WC10inA2HDM} \\[0.2cm]
&C_{7}^{\prime\mathrm{H^\pm}}=\frac{\bar m_{s}}{\bar m_{b}}\left(\left|\varsigma_{u}\right|^{2}C_{7,\,\mathrm{uu}} +\varsigma_{u}\varsigma_{d}^{\ast}C_{7,\,\mathrm{ud}}\right)\,,\label{eq:WC7pinA2HDM} \\[0.2cm]
&C_{9}^{\prime\mathrm{H^\pm}}=\left(-1+4\sin^{2}\theta_{W}\right)C_{10}^{\prime\mathrm{H^\pm}}+\frac{\bar m_{b}\bar m_{s}}{M_{W}^{2}}\Big[\left|\varsigma_{u}\right|^{2}C_{9,\,\mathrm{uu}}' +2\Re\left(\varsigma_{u}\varsigma_{d}^{\ast}\right)C_{9,\,\mathrm{ud}}' +\left|\varsigma_{d}\right|^{2}C_{9,\,\mathrm{dd}}'\Big]\,,\label{eq:WC9pinA2HDM} \\[0.2cm]
&C_{10}^{\prime\mathrm{H^\pm}}=\frac{\bar m_{b}\bar m_{s}}{M_{W}^{2}}\Big[\left|\varsigma_{u}\right|^{2}C_{10,\,\mathrm{uu}}' +2\Re\left(\varsigma_{u}\varsigma_{d}^{\ast}\right)C_{10,\,\mathrm{ud}}'+\left|\varsigma_{d}\right|^{2} C_{10,\,\mathrm{dd}}'\Big]\,,\label{eq:WC10pinA2HDM}
\end{align}
where the coefficients of the different combinations of the couplings $\varsigma_u$ and $\varsigma_d$ are given by Eqs.~(\ref{eq:WCA2HDM1})--(\ref{eq:WCA2HDM10}). In the particular cases of type II and type Y 2HDMs with large $\tan\beta$, the only terms enhanced by a factor $\tan^2\beta$ originate from the $\left|\varsigma_d\right|^2$ part contributing only to $C_{9,10}^{\prime\mathrm{H^\pm}}$. The Wilson coefficients $C_{7,9,10}^{(\prime)\mathrm{H^\pm}}$ are found to be invariant under a global U(1) transformation, $\varsigma_{u}\to\mathrm{e}^{i\chi}\varsigma_{u}$ and $\varsigma_{d}\to\mathrm{e}^{i\chi}\varsigma_{d}$. This invariance is well anticipated since it corresponds to an unphysical phase transformation of the second Higgs doublet, $\Phi_2\to \mathrm{e}^{i\chi}\Phi_2$, a leftover freedom in the Higgs basis~\cite{Davidson:2005cw,Haber:2006ue}. There is an implicit $\mu_W$ dependence via the $s,b,t$-quark masses, which depend on the precise definitions and have to be specified when going beyond the leading logarithm (LL). As we evaluate $C_{7,9,10}^{(\prime)\mathrm{H^\pm}}$ only at the leading order (LO) in $\alpha_s$, whether the running masses $\bar m_q(\mu_W)$ or the pole masses $m_q$ are used does not matter too much. As a consequence, we choose the pole masses $m_q$ as input in Eqs.~(\ref{eq:WC7pinA2HDM})--(\ref{eq:WC10pinA2HDM}).

Our results for the chirality-flipped Wilson coefficients $C_{7,9,10}^{\prime\mathrm{H^\pm}}$ are presented for the first time in the A2HDM. In the particular cases of the $\mathcal{Z}_2$ symmetric 2HDMs, our results agree with the ones calculated in refs.~\cite{Grinstein:1990tj,Bertolini:1990if,Cho:1996we,Chankowski:2000ng}. It is also noted that the next-to-leading order QCD corrections to $C_{7,9,10}^\mathrm{H^\pm}$ in the supersymmetry and type-II 2HDM have already been calculated in refs.~\cite{Ciuchini:1997xe,Borzumati:1998tg,Bobeth:1999ww,Bobeth:2001jm,Schilling:2004gk}.

\subsection{Angular observables in $\boldsymbol{B^0\to K^{\ast 0}\mu^+\mu^-}$ decay}

The angular distribution of the $B^0\to K^{\ast 0}(\to K^+\pi^-)\mu^+\mu^-$ decay is described by the dimuon invariant mass squared $q^2$ as well as the three angles $\theta_\ell$, $\theta_{K^\ast}$ and $\phi$, where $\theta_\ell$ is defined as the angle between the flight direction of the $\mu^+\,(\mu^-)$ and the opposite direction of the $B^0\,(\bar B^0)$ in the rest frame of the dimuon system, and $\theta_{K^\ast}$ the angle between the flight direction of the $K^+\,(K^-)$ and that of the $B^0\,(\bar B^0)$ in the $K^{\ast 0}\,(\bar K^{\ast 0})$ rest frame, while $\phi$ is the angle between the plane containing the dimuon pair and the plane containing $K^+$ and $\pi^-$ mesons in the $B^0\,(\bar B^0)$ rest frame. In terms of these four kinematic variables, the full angular decay distribution of the decay is then given by~\cite{Altmannshofer:2008dz,Kruger:1999xa}
\begin{align} \label{eq:angularlist}
  \frac{d^4 \bar{\Gamma}\left[B^0\to K^{\ast 0}\mu^+\mu^-\right]}{dq^2\, d\cos\theta_\ell\, d\cos\theta_{K^*}\, d\phi}& =\frac{9}{32\pi}\Big[
      \bar I_1^s \sin^2\theta_{K^\ast} + \bar I_1^c \cos^2\theta_{K^\ast}
      + (\bar I_2^s \sin^2\theta_{K^\ast} + \bar I_2^c \cos^2\theta_{K^\ast}) \cos 2\theta_\ell
\nonumber \\
    & \hspace{1.0cm} + \bar I_3 \sin^2\theta_{K^\ast} \sin^2\theta_\ell \cos 2\phi
      + \bar I_4 \sin 2\theta_{K^\ast} \sin 2\theta_\ell \cos\phi
\nonumber \\
    & \hspace{1.0cm} + \bar I_5 \sin 2\theta_{K^\ast} \sin\theta_\ell \cos\phi
\nonumber \\
    & \hspace{1.0cm} + \bar I_6^s \sin^2\theta_{K^\ast} \cos\theta_\ell
      + \bar I_7 \sin 2\theta_{K^\ast} \sin\theta_\ell \sin\phi
\nonumber \\
    & \hspace{1.0cm} + \bar I_8 \sin 2\theta_{K^\ast} \sin 2\theta_\ell \sin\phi
      + \bar I_9 \sin^2\theta_{K^\ast} \sin^2\theta_\ell \sin 2\phi\Big]\,,
\end{align}
where the angular coefficients $\bar I_i^{(a)}$ are functions of $q^2$ only, and the relations $\bar I_1^s=3\bar I_2^s$, $\bar I_1^c=-\bar I_2^c$ and $\bar I_6^c=0$ hold when the muon mass is neglected. The corresponding expression for the CP-conjugated mode $\bar B^0\to \bar K^{*0}(\to K^-\pi^+)\mu^+\mu^-$ is obtained from Eq.~(\ref{eq:angularlist}) by the replacements $\bar I_i^{(a)} \to I_i^{(a)}$~\cite{Altmannshofer:2008dz,Kruger:1999xa}. Explicit forms of the angular coefficients $\bar I_i^{(a)}~(I_i^{(a)})$ could be found, for example, in refs.~\cite{Altmannshofer:2008dz,Gratrex:2015hna,Aaij:2015oid}.

The self-tagging property of the decay $B^0\to K^{\ast 0}\mu^+\mu^-$ makes it possible to determine both the CP-averaged and the CP-asymmetric quantities defined, respectively, by~\cite{Altmannshofer:2008dz}
\begin{eqnarray} \label{eq:CP}
S_i^{(a)}=\left(I_i^{(a)}+\bar I_i^{(a)}\right)/\left(\frac{d\Gamma}{dq^2}+\frac{d\bar\Gamma}{dq^2}\right)\,, \qquad
A_i^{(a)}=\left(I_i^{(a)}-\bar I_i^{(a)}\right)/\left(\frac{d\Gamma}{dq^2}+\frac{d\bar\Gamma}{dq^2}\right).
\end{eqnarray}
The previously studied observables, such as the $q^2$ distributions of the forward-backward asymmetry $A_{FB}$ and the CP asymmetry $A_{CP}$, can be expressed in terms of these angular observables.

With the structure of the amplitudes at large recoil, it is possible to build clean observables whose sensitivity to the $B\to K^\ast$ transition form factors is suppressed by $\alpha_s$ or $\Lambda_{\rm QCD}/m_b$~\cite{Descotes-Genon:2013vna}. These include the so-called $P_i^{\prime}$ and $P_i$ observables defined by~\cite{Descotes-Genon:2013vna,Becirevic:2011bp,Matias:2012xw}
\begin{align}\label{eq:optimizeobs}
& P_1=\frac{S_3}{2S_2^s},\quad
  P_2=\frac{S_6^s}{8S_2^s},\quad
  P_3=\frac{S_9}{4S_2^s}\,,\\[0.2cm]
& P_4^{\prime}=\frac{S_4}{2\sqrt{-S_2^sS_2^c}},\quad
  P_5^{\prime}=\frac{S_5}{2\sqrt{-S_2^sS_2^c}},\quad P_6^{\prime}=\frac{S_7}{2\sqrt{-S_2^sS_2^c}},\quad P_8^{\prime}=\frac{S_8}{2\sqrt{-S_2^sS_2^c}}\,.
\end{align}
The numerical impact of charged-scalar contributions to some of these observables will be discussed in the next section.

\section{Numerical results and discussions}
\label{sec:results}

\subsection{Choice of the model parameters}

For the considered decay $B^0\to K^{\ast 0}\mu^+\mu^-$, only three model parameters, the charged-scalar mass $M_{H^\pm}$ and the two alignment parameters $\varsigma_u$ and $\varsigma_d$, are involved. In the following we assume the parameters $\varsigma_{u,d}$ to be real, indicating that the only source of CP violation in the A2HDM is still due to the CKM matrix. Following the previous studies, we give below the preset ranges of these model parameters:
\begin{itemize}
  \item The charged-scalar mass is assumed to lie in the range $M_{H^{\pm}} \in[80,1000]~\mathrm{GeV}$, where the lower bound comes from the LEP direct search~\cite{Searches:2001ac}, while the upper bound from the unitarity and stability of the scalar potential~\cite{Barroso:2013awa,Dev:2014yca,Das:2015qva,Chakraborty:2015raa}.

  \item The alignment parameter $\varsigma_u$ is assumed to lie in the range $|\varsigma_u|\leq 2$, to be compatible with the current data of loop-induced processes, such as $Z\to b\bar{b}$, $b\to s \gamma$, $B_{s,d}^0-\bar{B}_{s,d}^0$ mixings, as well as the $h(125)$ decays~\cite{Jung:2010ab,Jung:2012vu,Celis:2012dk,Duarte:2013zfa,Jung:2013hka,Celis:2013ixa,Celis:2013rcs}.

  \item The alignment parameter $\varsigma_d$ is only mildly constrained through phenomenological requirements that involve additionally other model parameters. So we let it to be a free parameter.

  \item In the 2HDMs with discrete $\mathcal{Z}_2$ symmetries, the parameters $\varsigma_{u}$ and $\varsigma_{d}$ are not independent but are related to each other through the ratio of the VEVs $\tan\beta=v_2/v_1$. The upper limit for $\tan\beta$ also comes from the unitarity and stability of the scalar potential~\cite{Barroso:2013awa,Dev:2014yca,Das:2015qva,Chakraborty:2015raa}; we assume here $\tan\beta\leq50$.
\end{itemize}

\subsection{Constraints on the model parameters}

For the other input parameters, we take $M_Z=91.1876~\mathrm{GeV}$, $M_W=80.385~\mathrm{GeV}$, $m_t=(174.2\pm1.4)~\mathrm{GeV}$, $m_b=(4.78\pm0.06)~\mathrm{GeV}$, and $\bar m_s(2~\mathrm{GeV})=(96^{+8}_{-4})~\mathrm{MeV}$~\cite{Olive:2016xmw}. Since $C_7^{\prime \mathrm{H}^\pm}=\bar{m}_s/\bar{m}_b C_7^{\mathrm{H}^\pm}$ and $\bar m_s \ll \bar m_b$, the contribution from $O_7'$ will be safely neglected.

The Wilson coefficient $C_7^{\mathrm{H}^\pm}$ is severely constrained by the inclusive decay $B\to X_s \gamma$. The branching ratio of $B\to X_s \gamma$ measured by CLEO~\cite{Chen:2001fja}, Belle~\cite{Limosani:2009qg,Saito:2014das,Aubert:2007my} and BaBar~\cite{Lees:2012ym,Lees:2012ufa,Lees:2012wg}, lead to the combined average~\cite{Amhis:2014hma}
\begin{eqnarray}\label{eq:exp_b2sgamma}
\mathcal{B}^{\mathrm{exp}}(B\to X_s \gamma)\big|_{E_\gamma>1.6~\mathrm{GeV}}\, =\,\left(3.43\pm0.21_{\rm stat.}\pm0.07_{\rm syst.}\right)\times10^{-4}\,,
\end{eqnarray}
which is in good agreement with the updated SM prediction~\cite{Misiak:2015xwa}
\begin{eqnarray}\label{eq:sm_b2sgamma}
\mathcal{B}^{\mathrm{SM}}(B\to X_s \gamma)\big|_{E_\gamma>1.6~\mathrm{GeV}}\, =\,\left(3.36\pm0.23\right)\times10^{-4}.
\end{eqnarray}
It should be noted that the chromomagnetic dipole operator $O_8 = \frac{g_s}{16\pi^2}\bar m_b\left(\bar s\sigma^{\mu\nu}P_{R}T^a b\right)G_{\mu\nu}^a$ also plays an important role in the decay $B\to X_s \gamma$. However, at the LO in $\alpha_s$, this operator contributes to $B\to X_s \gamma$ only via its mixing with $O_7$. It is then found that, at the matching scale $\mu_W=160~\mathrm{GeV}$, the Wilson coefficients $C_7^{\mathrm{H}^\pm}$ and $C_8^{\mathrm{H}^\pm}$ should fulfill the constraint~\cite{Misiak:2015xwa}:
\begin{eqnarray}\label{eq:onlyC7}
-0.0634\leq C_7^{\mathrm{H}^\pm}\left(\mu_W\right)+0.242\,C_8^{\mathrm{H}^\pm}\left(\mu_W\right)\leq 0.0464\,,
\end{eqnarray}
where $C_8^{\mathrm{H}^\pm}=\left|\varsigma_{u}\right|^{2}C_{8,\,\mathrm{uu}}+\varsigma_{d}\varsigma_{u}^{\ast} C_{8,\,\mathrm{ud}}$~\cite{Grinstein:1990tj}, with the functions $C_{8,\,\mathrm{uu}}$ and $C_{8,\,\mathrm{ud}}$ given, respectively, by Eqs.~(\ref{eq:WCA2HDM11}) and (\ref{eq:WCA2HDM12}).

\begin{figure}[t]
  \centering
  \includegraphics[width=0.5\textwidth]{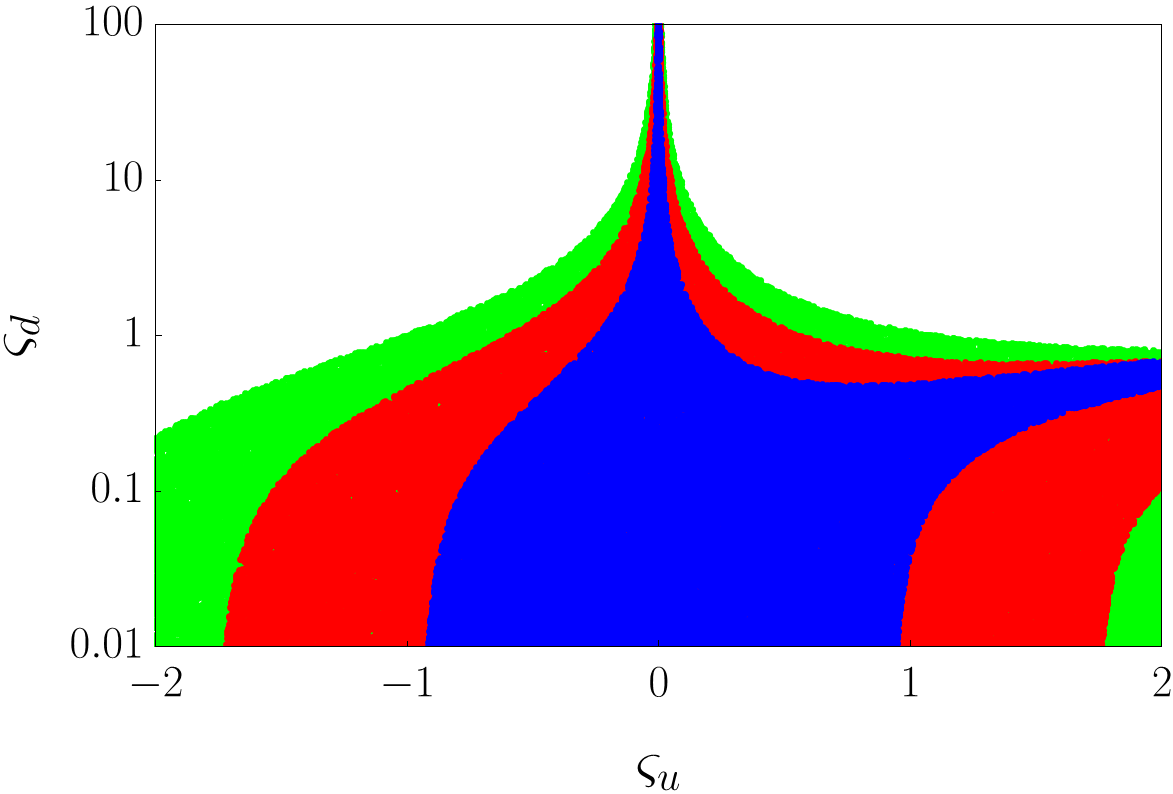}
  \caption{\small The allowed regions in the $\varsigma_u-\varsigma_d$ plane~($\varsigma_d>0$) under the constraint from Eq.~(\ref{eq:onlyC7}). The \textit{blue}, \textit{red}, and \textit{green bands} correspond to $M_{H^\pm}=80$, $300$ and $500~\mathrm{GeV}$, respectively.}
  \label{fig:reC7log}
\end{figure}

Under the constraint from Eq.~(\ref{eq:onlyC7}), we show in Figure~\ref{fig:reC7log} the allowed regions in the $\varsigma_u-\varsigma_d$ plane~($\varsigma_d>0$), with three representative values of the charged-scalar mass, $M_{H^\pm}=80$, $300$ and $500~\mathrm{GeV}$ as benchmarks. The case with $\varsigma_d<0$ is obtained from Figure~\ref{fig:reC7log} with the changes $\varsigma_u\to-\varsigma_u$ and $\varsigma_d\to-\varsigma_d$. It is observed that the allowed range of $\varsigma_d$ becomes quite large when $\varsigma_u$ tends to zero; particularly, when $\varsigma_u=0$, no constraint on $\varsigma_d$ is obtained, because in this limit the SM result is recovered. When $\varsigma_d=0$, on the other hand, a bound on $\varsigma_u$ can be set with the allowed range of $|\varsigma_u|$ further strengthened for smaller values of the charged-scalar mass. These qualitative observations are consistent with those observed previously in refs.~\cite{Jung:2010ab,Jung:2010ik,Jung:2012vu}. However, the allowed regions for $\varsigma_u$ and $\varsigma_d$ are further reduced compared to those obtained in refs.~\cite{Jung:2010ab,Jung:2010ik,Jung:2012vu}, because the updated SM prediction (cf. Eq.~\eqref{eq:sm_b2sgamma}) becomes now more compatible with the current experimental data (cf. Eq.~\eqref{eq:exp_b2sgamma}). It is also found that the preset maximum value $\left|\varsigma_u\right|=2$ is reached when $\left|\varsigma_d\right|$ varies within a range away from zero, rather than at $\varsigma_d=0$; for example, taking $M_{H^\pm}=80~\mathrm{GeV}$, we find that $\left|\varsigma_u\right|$ approaches to 2 when $0.6<\left|\varsigma_d\right|<0.8$. This novel observation motivates us to display the $\varsigma_d$-axis in the logarithmic coordinate, making clear the correlation between $\varsigma_u$ and $\varsigma_d$ in the range $\left|\varsigma_d\right|<1$. The inversely-proportional and parabolic boundary curves in the first quadrant indicate that the NP contribution to $C_{7}^\mathrm{H^\pm}$ (cf. Eq.~\eqref{eq:WC7inA2HDM}) is dominated by the $\varsigma_d\varsigma_u^\ast$ and $\left|\varsigma_u\right|^2$ terms, respectively. As the large same-sign solutions for $\varsigma_u$ and $\varsigma_d$ obtained in refs.~\cite{Jung:2010ab,Jung:2010ik}, corresponding to the case when the NP influence is about twice the size of the SM
contribution but with an opposite sign, are already excluded by the isospin asymmetry of $B\to K^\ast\gamma$ decays~\cite{Jung:2012vu,Li:2013vlx}, they are not shown in Figure~\ref{fig:reC7log}.

\begin{figure}[t]
  \centering
  \includegraphics[width=0.5\textwidth]{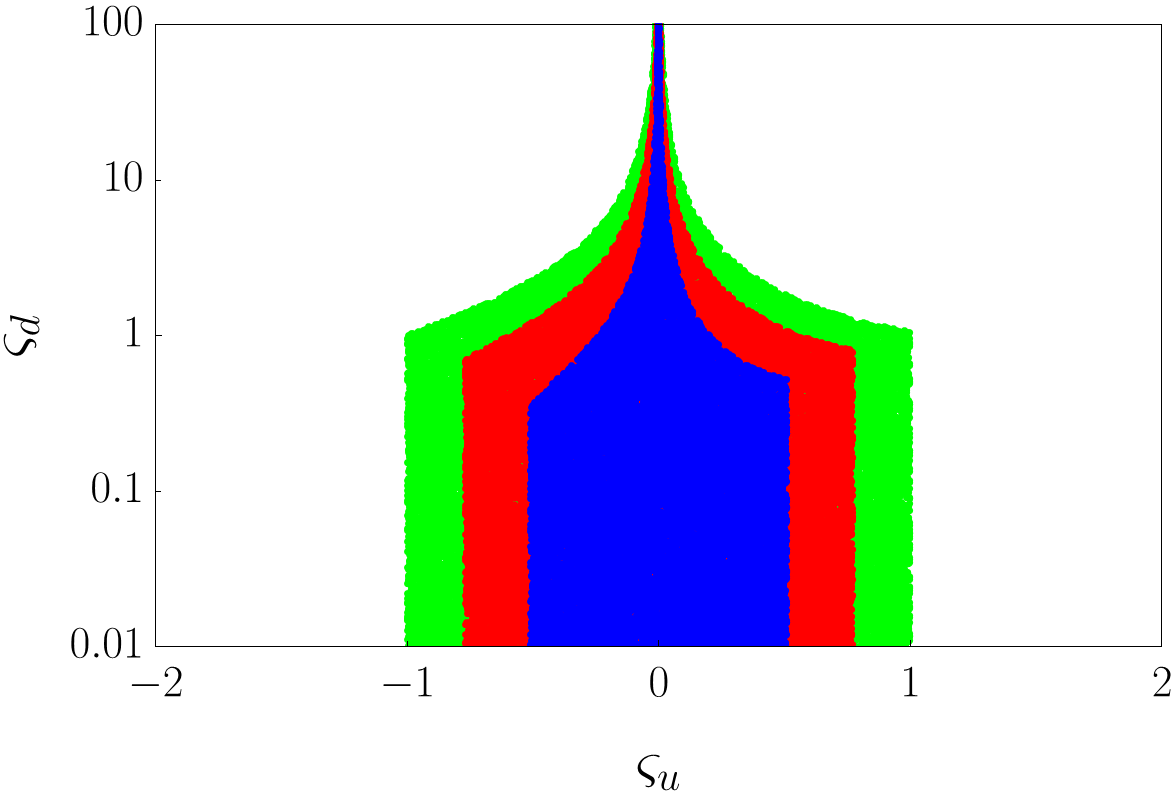}
  \caption{\small The allowed regions in the $\varsigma_u-\varsigma_d$ plane~($\varsigma_d>0$) under the constraint from Eq.~(\ref{eq:onlyC7}) as well as the bounds on $C_{9,10}^{\mathrm{H^\pm}}$ and $C_{9,10}^{\prime\mathrm{H^\pm}}$ from Eqs.~(\ref{uneq:WCJHEP}) and (\ref{uneq:WCPRD}). The other captions are the same as in Figure~\ref{fig:reC7log}.}
  \label{fig:reClog}
\end{figure}

Motivated by the latest LHCb and Belle measurements of $b \to s \ell^+\ell^-$ decays, there exist several global fits for the NP contributions to the Wilson coefficients $C_{9,10}^{(\prime)}$~\cite{Altmannshofer:2014rta,Descotes-Genon:2015uva,Hurth:2016fbr,Meinel:2016grj}. We use two of these global fit results to further constrain the A2HDM parameters. One is obtained from the combined fit to the $b\to s\, (\mu^+\mu^-, \gamma)$ mesonic decays (at $\mu_b=4.8~\mathrm{GeV}$)~\cite{Descotes-Genon:2015uva}:
\begin{align} \label{uneq:WCJHEP}
-2.2\,\leq\, &C_9^{\mathrm{NP}}\,\leq\,-0.4\,, & -0.5\,\leq\, &C_{10}^{\mathrm{NP}}\,\leq\,2.0\,,\nonumber\\[0.2cm]
-1.3\,\leq\, &C_9^{\prime\mathrm{NP}}\,\leq\,3.7\,, & -1.0\,\leq\, &C_{10}^{\prime\mathrm{NP}}\,\leq\,1.6\,,
\end{align}
given at the $3\sigma$ level. This fit includes the branching ratios and optimized angular observables of $B\to K^\ast\mu^+\mu^-$ and $B_s\to\phi\mu^+\mu^-$, the branching ratios of $B\to K\mu^+\mu^-$, the branching ratios of $B\to X_s\mu^+\mu^-$ (restricted only to the range $1~\mathrm{GeV}^2\leq q^2\leq 6~\mathrm{GeV}^2$) and $B\to X_s\gamma$, the branching ratio of $B_s\to\mu^+\mu^-$, as well as the isospin asymmetry and the time-dependent CP asymmetry of $B\to K^\ast\gamma$. Furthermore, both the large- and low-recoil data is included for the exclusive $b\to s\mu^+\mu^-$ decays, resulting in nearly a hundred observables in total in the analysis~\cite{Descotes-Genon:2015uva}. The other global fit includes, besides the time-integrated branching ratio of $B_s\to\mu^+\mu^-$ and the branching ratio of $B\to X_s\ell^+\ell^-$ integrated over the range $1~\mathrm{GeV}^2\leq q^2\leq 6~\mathrm{GeV}^2$, the currently available data on $\Lambda_b\to\Lambda (\to p\pi^-)\,\mu^+\mu^-$ decay, which involves the branching ratio, the rate of longitudinally polarized lepton pair, as well as the leptonic and the hadronic forward-backward asymmetries; numerically, this fit gives (at $\mu_b=4.2~\mathrm{GeV}$)~\cite{Meinel:2016grj}:
\begin{align} \label{uneq:WCPRD}
0.9\,\leq\, &C_9^{\mathrm{NP}}\,\leq\,2.5\,, & 1.8\,\leq\, &C_{10}^{\mathrm{NP}}\,\leq\,4.2\,,\nonumber\\[0.2cm]
-1.3\,\leq\, &C_9^{\prime\mathrm{NP}}\,\leq\,1.8\,, & 1.0\,\leq\, &C_{10}^{\prime\mathrm{NP}}\,\leq\,3.1\,,
\end{align}
at the $1\sigma$ level. It is interesting to note that the latter prefers a shift to $C_9$ that is opposite in sign compared to the former~\cite{Meinel:2016grj}. Since the Wilson coefficients $C_{9,10}^{\mathrm{H^\pm}}(\mu_W)$ and $C_{9,10}^{\prime\mathrm{H^\pm}}(\mu_W)$ are calculated only at the LO, they should be evolved to the lower scale $\mu_b$ at the LL approximation, which means that they are actually not running~\cite{Gambino:2003zm}. Thus, we can apply directly the bounds given by Eqs.~(\ref{uneq:WCJHEP}) and (\ref{uneq:WCPRD}) to $C_{9,10}^{\mathrm{H^\pm}}$ and $C_{9,10}^{\prime\mathrm{H^\pm}}$. To be more conservative, we require each of these coefficients to lie within the smaller lower and bigger upper bounds of these two global fits. Using these bounds as well as the constraint from Eq.~(\ref{eq:onlyC7}), we find that the allowed parameter space in the $\varsigma_u-\varsigma_d$ plane are significantly reduced, especially for the model parameter $\varsigma_u$, as shown in Figure~\ref{fig:reClog}. This means that $C_{9,10}^\mathrm{H^\pm}$ play a major role in the small $\left|\varsigma_d\right|$ region~($\left|\varsigma_d\right|<1$) and $C_{9,10}^{\prime\mathrm{H^\pm}}$ can be quite sizable when $\varsigma_u$ approaches to zero.

\begin{figure}[t]
  \centering
  \includegraphics[width=0.9\textwidth]{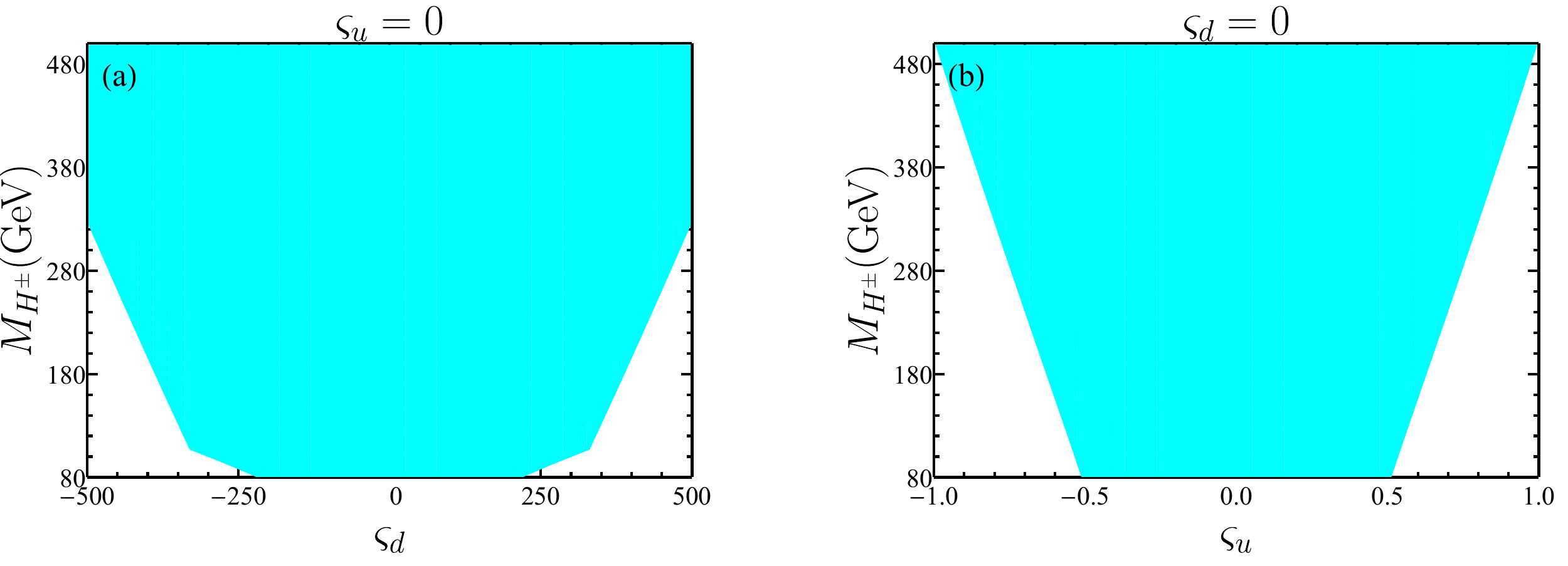}
  \caption{\small The allowed regions in the $\varsigma_d-M_{H^\pm}$ plane when $\varsigma_u=0$ (\textbf{a}) and in the $\varsigma_u-M_{H^\pm}$ plane when $\varsigma_d=0$ (\textbf{b}), under the constraint from Eq.~(\ref{eq:onlyC7}) as well as the bounds on $C_{9,10}^{\mathrm{H^\pm}}$ and $C_{9,10}^{\prime\mathrm{H^\pm}}$ from Eqs.~(\ref{uneq:WCJHEP}) and (\ref{uneq:WCPRD}).}
  \label{fig:reZero}
\end{figure}

It is also interesting to note that, under the constraint from Eq.~(\ref{eq:onlyC7}) as well as the bounds on $C_{9,10}^{\mathrm{H^\pm}}$ and $C_{9,10}^{\prime\mathrm{H^\pm}}$ from Eqs.~(\ref{uneq:WCJHEP}) and (\ref{uneq:WCPRD}), we could obtain a bound on $\varsigma_d$ even when $\varsigma_u$ equals to zero. Such a bound arises entirely from the information on $C_{9,10}^{\prime\mathrm{H^\pm}}$ due to the $\left|\varsigma_d\right|^2$ terms in these two Wilson coefficients (cf. Eqs.~\eqref{eq:WC9pinA2HDM} and \eqref{eq:WC10pinA2HDM}). For illustration, the allowed regions in the $\varsigma_d-M_{H^\pm}$ plane when $\varsigma_u=0$ and in the $\varsigma_u-M_{H^\pm}$ plane when $\varsigma_d=0$ are shown in Figure~\ref{fig:reZero}. Numerically, we obtain $\left|\varsigma_u\right|\leq 0.506$, $0.763$ and $0.990$, and $\left|\varsigma_d\right|\leq 212$, $476$ and $622$,  corresponding to $M_{H^\pm}=80$, $300$ and $500~\mathrm{GeV}$, respectively. This means that the more accurate $C_{9,10}^{\prime{\rm NP}}$ can be better used to restrict the parameter $\varsigma_d$.

\subsection{$P_2$ and $P_5'$ in the A2HDM}

\begin{figure}[htbp]
  \centering
  \includegraphics[width=0.80\textwidth]{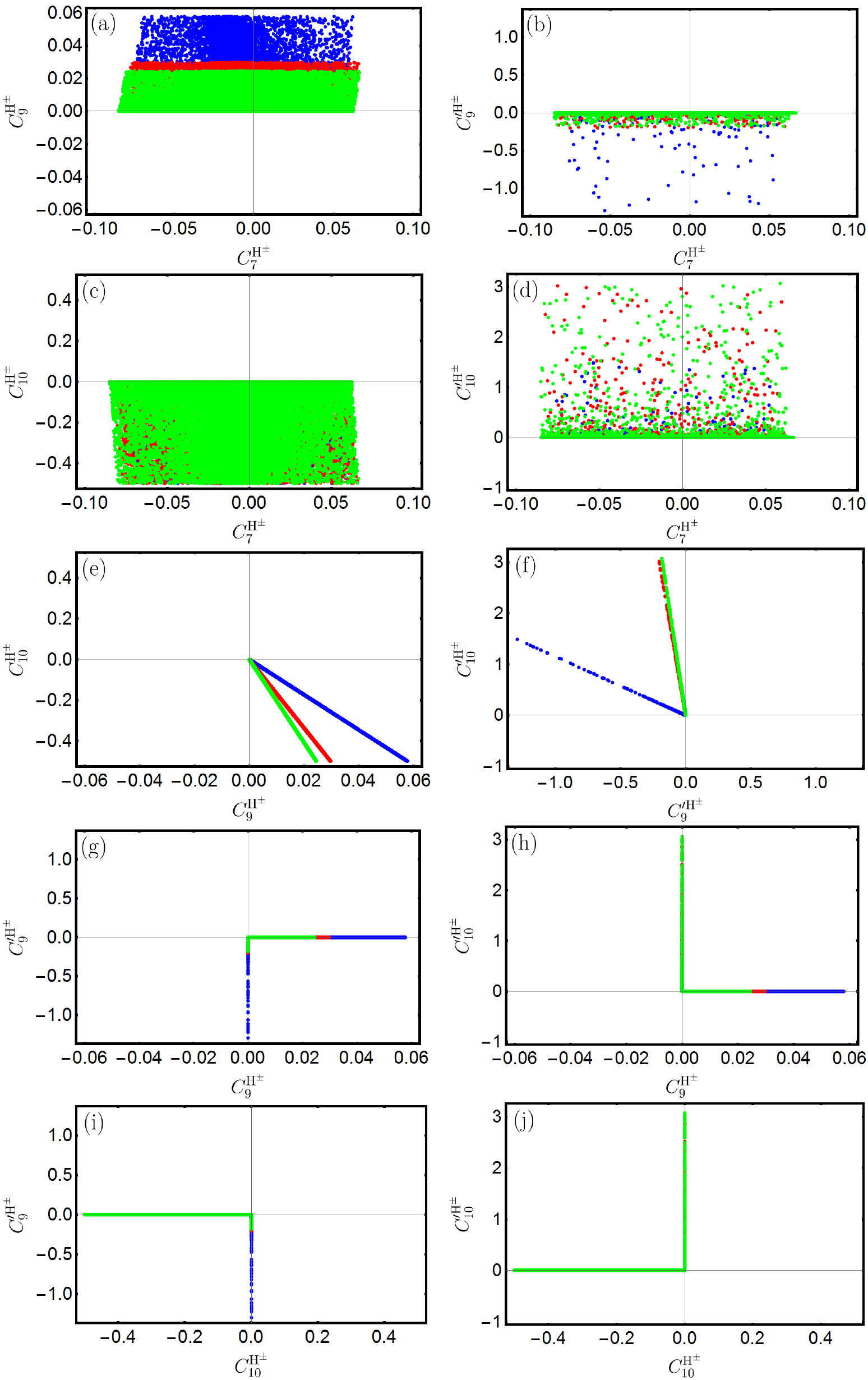}
  \caption{\small Correlations among the five Wilson coefficients using the allowed values of $\varsigma_u$ and $\varsigma_d$ with three benchmark values of charged-scalar mass obtained in the previous subsection. The other captions are the same as in Figure~\ref{fig:reC7log}.}
  \label{fig:wcicorr}
\end{figure}

In this subsection, with the constrained parameter space for $\varsigma_u$ and $\varsigma_d$, we investigate the impact of A2HDM on the angular observables $P_2$ and $P_5'$ in the decay $B^0\to K^{\ast 0}\mu^+\mu^-$. As there involve only three model parameters $\varsigma_u$, $\varsigma_d$ and $M_{H^\pm}$ in Eqs.~(\ref{eq:WC7inA2HDM})--(\ref{eq:WC10pinA2HDM}), the five Wilson coefficients ($C_7^{\prime\mathrm{H^\pm}}$ is neglected because $\bar m_s \ll \bar m_b$) are expected to be highly correlated with each other. Using the allowed values of $\varsigma_u$ and $\varsigma_d$ with three benchmark values of charged-scalar mass obtained in the previous subsection, we show in Figure~\ref{fig:wcicorr} the correlations among these five Wilson coefficients. One can see that, while $C_7^\mathrm{H^\pm}$ is hardly correlated with the other four Wilson coefficients (Figures~\ref{fig:wcicorr}(a)--\ref{fig:wcicorr}(d)), $C_9^\mathrm{H^\pm}$ and $C_{10}^\mathrm{H^\pm}$ are obviously linearly correlated with each other and the slope depends only on the charged-scalar mass $M_{H^\pm}$ (Figure~\ref{fig:wcicorr}(e)), with the blue, red, and green lines obtained with $M_{H^\pm}=80$, $300$, and $500~\mathrm{GeV}$, respectively. In addition, $C_9^{\prime\mathrm{H^\pm}}$ and $C_{10}^{\prime\mathrm{H^\pm}}$ are found to be approximately linearly correlated with each other (Figure~\ref{fig:wcicorr}(f)), and the slope starts to be nearly a constant when $M_{H^\pm}\geq250~\mathrm{GeV}$, which explains why the two lines with $M_{H^\pm}=300$ and $500~\mathrm{GeV}$ almost overlap completely in Figure~\ref{fig:wcicorr}(f). In fact, from the analytic expressions for these Wilson coefficients (cf. Eqs.~(\ref{eq:WC9inA2HDM})--(\ref{eq:WC10inA2HDM}) and (\ref{eq:WC9pinA2HDM})--(\ref{eq:WC10pinA2HDM}), together with (\ref{eq:WCA2HDM3})--(\ref{eq:WCA2HDM10})), we find that $C_{9}^\mathrm{H^\pm}/C_{10}^\mathrm{H^\pm}\to -1+4\sin^{2}\theta_{W}\,\left[1+4/(9x_t)\right]$ and $C_{9}^{\prime\mathrm{H^\pm}}/C_{10}^{\prime\mathrm{H^\pm}}\to -1+4\sin^{2}\theta_{W}$ when $M_{H^\pm}$ goes to infinity. This explains why the lines shown in Figures~\ref{fig:wcicorr}(e) and \ref{fig:wcicorr}(f) get closer to each other with larger $M_{H^\pm}$.

The most interesting results are shown in Figures~\ref{fig:wcicorr}(g)--\ref{fig:wcicorr}(j), which suggest that the charged scalars can not affect the left- and right-handed semileptonic operators at the same time, under the constraints shown in Figures~\ref{fig:reC7log} and \ref{fig:reClog}. According to Eqs.~(\ref{eq:WC9inA2HDM}) and (\ref{eq:WC10inA2HDM}), sizable $C_{9,10}^\mathrm{H^\pm}$ need a large $\left|\varsigma_u\right|$, which in turn implies that $\left|\varsigma_d\right|$ can not be too large due to the constraints shown in Figures~\ref{fig:reC7log} and \ref{fig:reClog}. Together with the small factor $\bar m_b\bar m_s/M_W^2$ and the preset range $\left|\varsigma_u\right|\leq2$, this renders the coefficients $C_{9,10}^{\prime\mathrm{H^\pm}}$ quite small (cf. Eqs.~(\ref{eq:WC9pinA2HDM})--(\ref{eq:WC10pinA2HDM})). The same argument applies to the opposite case: sizable $C_{9,10}^{\prime\mathrm{H^\pm}}$ are possible only with a large $\left|\varsigma_d\right|$, which then implies a small $\left|\varsigma_u\right|$, resulting in quite small $C_{9,10}^\mathrm{H^\pm}$. These qualitative analyses explain the strong correlations observed in Figures~\ref{fig:wcicorr}(g)--\ref{fig:wcicorr}(j), and motivate us to consider the following two specific cases for the NP Wilson coefficients:
\begin{align}
&\text{Case A: $C_{7,9,10}^\mathrm{H^\pm}$ are sizable, but $C_{9,10}^{\prime\mathrm{H^\pm}}\simeq0$;} \\[0.2cm]
&\text{Case B: $C_7^\mathrm{H^\pm}$ and $C_{9,10}^{\prime\mathrm{H^\pm}}$ are sizable, but $C_{9,10}^\mathrm{H^\pm}\simeq0$.}
\end{align}
They are associated to the (large $\left|\varsigma_u\right|$, small $\left|\varsigma_d\right|$) and (small $\left|\varsigma_u\right|$, large $\left|\varsigma_d\right|$) regions, respectively.

\begin{figure}[t]
  \centering
  \includegraphics[width=0.9\textwidth]{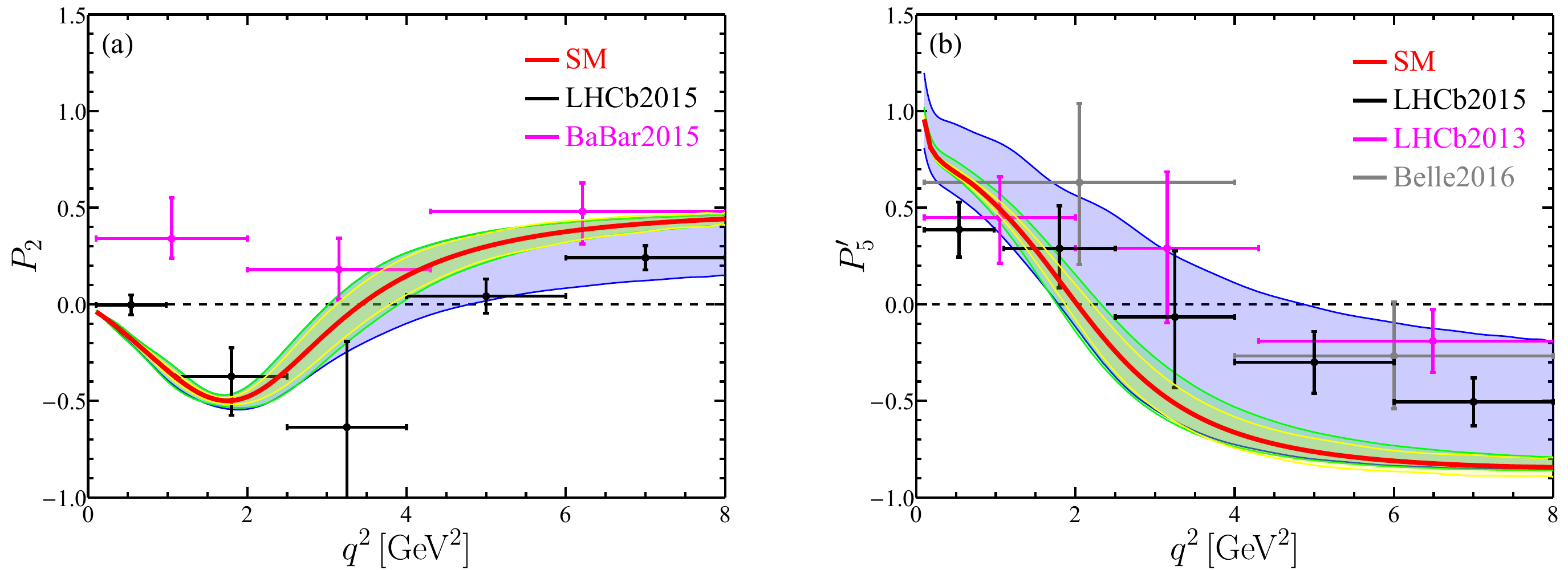}
  \caption{\small The $q^2$ dependence of the angular observables $P_2$ and $P_5'$, both within the SM (central value by a \textit{red curve} and its uncertainty by a \textit{yellow band}) and in the A2HDM (the \textit{green} and \textit{blue bands} correspond to the case~A and case~B, respectively). The experimental data from the LHCb~\cite{Aaij:2013qta,Aaij:2015oid}, Belle~\cite{Abdesselam:2016llu} and BaBar~\cite{Lees:2015ymt} collaborations are represented by the corresponding \textit{error bars} in different $q^2$ bins.}
  \label{fig:resultP25}
\end{figure}

In Figure~\ref{fig:resultP25}, we show our predictions for the two angular observables $P_2$ and $P_5'$ at large recoil both within the SM and in the A2HDM, with the Wilson coefficients obtained in the above two cases, together with the experimental data from the LHCb~\cite{Aaij:2013qta,Aaij:2015oid}, Belle~\cite{Abdesselam:2016llu} and BaBar~\cite{Lees:2015ymt} collaborations. Here we follow closely the method used in refs.~\cite{Altmannshofer:2008dz,Altmannshofer:2014rta,Straub:2015ica}: Firstly, we take as input the combined LCSR-lattice fit results for the $B\to K^\ast$ transition form factors provided in ref.~\cite{Straub:2015ica}, which allow us to retain all the correlated uncertainties among these form factors. Secondly, we have included the hadronic uncertainties due to non-factorizable power corrections associated with the non-perturbative charm loops~\cite{Khodjamirian:2010vf,Straub:2015ica}, the latest discussions of which could be found in refs.~\cite{Capdevila:2017ert,Chobanova:2017ghn}. Finally, these two angular observables are computed within the SM, with their respective uncertainties obtained by adding in quadrature the individual uncertainty due to the $B\to K^\ast$ form factors, the non-factorizable charm-loop contributions, and the parametric input (mainly from $\bar m_b(\bar m_b)=4.18^{+0.04}_{-0.03}\,{\rm GeV}$ and $m_c=1.4\pm0.2\,{\rm GeV}$). For the NP contributions, however, we consider only the uncertainties of the model parameters and perform a random flat scan within their allowed regions. One can see clearly that there is only a small impact on $P_2$ and $P_5'$ in case A, where the chirality-flipped operators $O'_{9,10}$ are absent, while in case B $P_5'$ could be increased significantly to be consistent with the experimental data and reduce $P_2$ when the dimuon invariant mass squared $q^2$ is higher than the zero-crossing point $q_0^2$. Numerical results for the zero-crossing points of $P_2$ (nonzero one) and $P_5'$ are given in Table~\ref{tab:zero}, both within the SM and in the A2HDM. It is observed that the impact on $q_0^2$ in case B is more pronounced than in case A.

\begin{table}[t]
\tabcolsep 0.2in
\renewcommand\arraystretch{1.3}
\begin{center}
\caption{\small \label{tab:zero} The zero-crossing points of $P_2$ (nonzero one) and $P_5'$ both within the SM and in the A2HDM.}\vspace{0.2cm}
\begin{tabular}{|c|c|c|c|}
\hline
        & SM & Case~A & Case~B  \\
\hline
$q^2_0(P_2)$ & $3.43^{+0.33}_{-0.32}$ & $(3.02,\,3.90)$  & $(3.02,\,4.79)$ \\
\hline
$q^2_0(P_5')$ & $2.02^{+0.19}_{-0.15}$ & $(1.77,\,2.32)$  & $(1.79,\,4.85)$  \\
\hline
\end{tabular}
\end{center}
\end{table}

\subsection{2HDMs with $\mathcal{Z}_2$ symmetries}

\begin{figure}[t]
  \centering
  \includegraphics[width=0.5\textwidth]{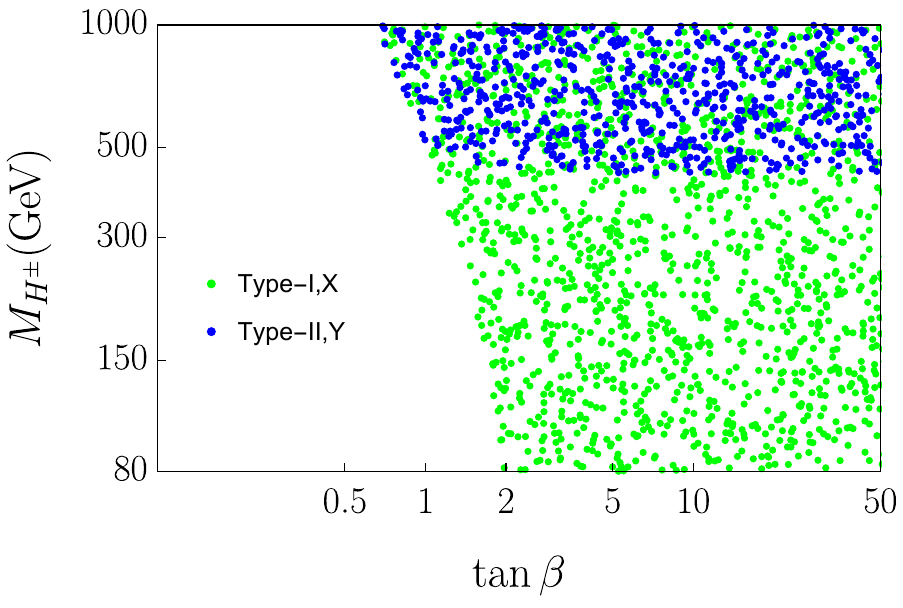}
  \caption{\small Allowed regions in the $\tan\beta-M_{H^\pm}$ plane corresponding to different $\mathcal{Z}_2$-symmetric 2HDMs, under the constraint from Eq.~(\ref{eq:onlyC7}) as well as the bounds on $C_{9,10}^{\mathrm{H^\pm}}$ and $C_{9,10}^{\prime\mathrm{H^\pm}}$ from Eqs.~(\ref{uneq:WCJHEP}) and (\ref{uneq:WCPRD}).}
  \label{fig:reZ2}
\end{figure}

In the generic 2HDMs with discrete $\mathcal{Z}_2$ symmetries, the three alignment parameters $\varsigma_f$ will be reduced to a single parameter $\tan\beta=v_2/v_1\geq0$, as indicated in Table~\ref{tab:models}. There are, therefore, only two model parameters $\tan\beta$ and $M_{H^\pm}$ in the Wilson coefficients $C_{7,9,10}^{(\prime)\mathrm{H^\pm}}$. We show in Figure~\ref{fig:reZ2} the allowed regions in the $\tan\beta-M_{H^\pm}$ plane corresponding to the four different types of 2HDMs with $\mathcal{Z}_2$ symmetries. As $C_{7,9,10}^{(\prime)\mathrm{H^\pm}}$ do not depend on the parameter $\varsigma_\ell$, the type I (II) and type X (Y) models are indistinguishable from each other. However, one can clearly distinguish types I and X from types II and Y models. As shown in Figure~\ref{fig:reZ2}, the bound $M_{H^\pm}>432\,{\rm GeV}$ is obtained for types II and Y 2HDMs, while there is no further bound found for $M_{H^\pm}$ in types I and X 2HDMs with sizable $\tan\beta$.

\begin{figure}[t]
  \centering
  \includegraphics[width=0.9\textwidth]{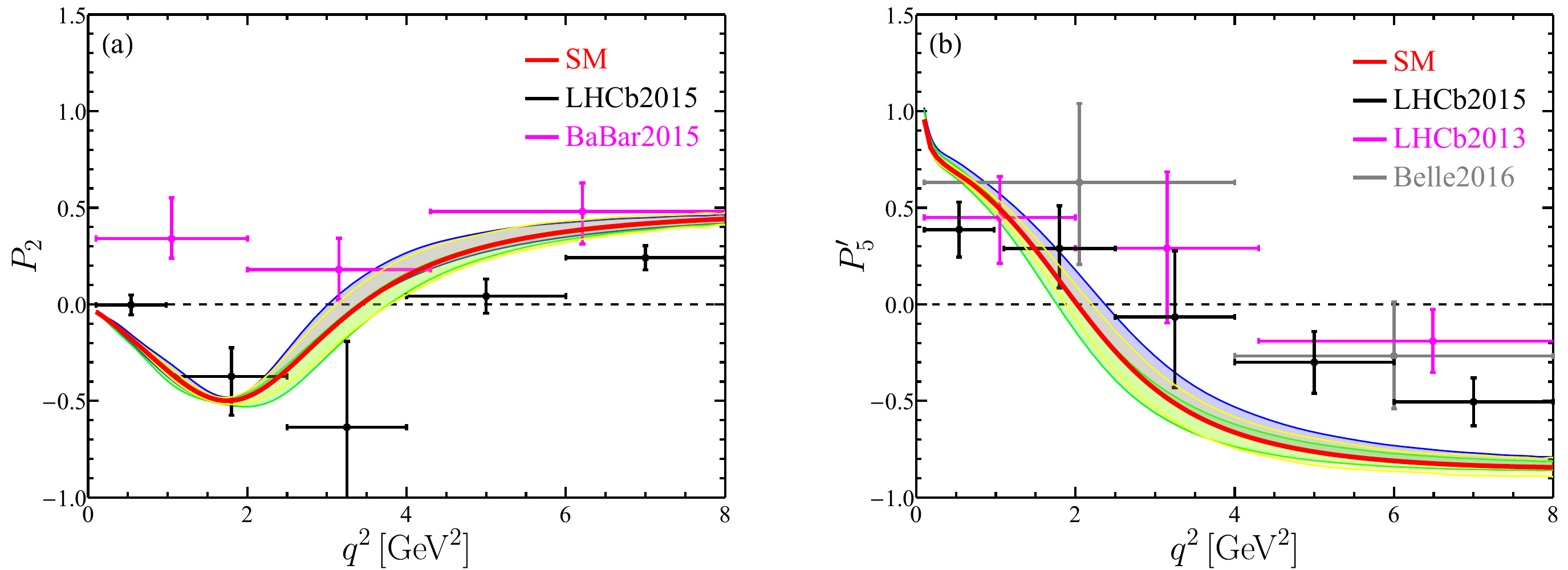}
  \caption{\small The $q^2$ dependence of the angular observables $P_2$ and $P_5'$ in the types I and X (the \textit{green band}) and the types II and Y (the blue band) 2HDMs. The other captions are the same as in Figure~\ref{fig:resultP25}.}
  \label{fig:NFCP25}
\end{figure}

With the constrained model parameters shown in Figures~\ref{fig:reZ2}, we then show in Figure~\ref{fig:NFCP25} the $q^2$ dependence of $P_2$ and $P_5'$ in the four different types of 2HDMs with $\mathcal{Z}_2$ symmetries. One can see that, compared to the SM predictions, both $P_2$ and $P_5'$ are reduced in the types I and X (the green band), but increased in the types II and Y (the blue band) 2HDMs, only by a small amount. This is because the charged-scalar effect on the left- and right-handed semileptonic operators is controlled by the same parameter $\tan\beta$ and, under the constraint shown in Figures~\ref{fig:reZ2}, sizable $C_{9,10}^{\prime\mathrm{H^\pm}}$ are not allowed in these models. It is, therefore, concluded the 2HDMs with $\mathcal{Z}_2$ symmetries can not explain the so-called $P_5'$ anomaly.

\section{Conclusions}
\label{sec:conclusion}

In this paper, we have presented a complete one-loop calculation of the SD Wilson coefficients $C_{7,9,10}^{(\prime)\mathrm{H^\pm}}$ due to the charged-scalar exchanges through the $Z^0$- and $\gamma$-penguin diagrams within the A2HDM. For $C_{9,10}^{\prime\mathrm{H^\pm}}$, although being suppressed by the factor $\bar m_b\,\bar m_s/M_W^2$, they could play an important role in interpreting the observed $P_5'$ anomaly in the decay $B^0\to K^{\ast 0}\mu^+\mu^-$, when the model parameter $|\varsigma_d|$ is large.

Under the constraints from the branching ratio $\mathcal{B}(B\to X_s\gamma)$ and the recent global fit results of $b\to s\ell^+\ell^-$ data, we have obtained the allowed parameter spaces in the $\varsigma_u-\varsigma_d$ plane, corresponding to three representative charged-scalar masses. We found that $C_{9,10}^\mathrm{H^\pm}$ play a major role in the small $\left|\varsigma_d\right|$ region ($\left|\varsigma_d\right|<1$), while $C_{9,10}^{\prime\mathrm{H^\pm}}$ are most important when the model parameter $\varsigma_u$ approaches to zero. When $\varsigma_u$ is far away from zero and $\left|\varsigma_d\right|\geq1$, on the other hand, the impact of $C_{7}^\mathrm{H^\pm}$ will become more significant. Within the constrained parameter space, numerically, the effects of these NP Wilson coefficients can be divided into the following two cases: (A) $C_{7,9,10}^\mathrm{H^\pm}$ are sizable, but $C_{9,10}^{\prime\mathrm{H^\pm}}\simeq0$, corresponding to the (large $\left|\varsigma_u\right|$, small $\left|\varsigma_d\right|$) region; (B) $C_7^\mathrm{H^\pm}$ and $C_{9,10}^{\prime\mathrm{H^\pm}}$ are sizable, but $C_{9,10}^\mathrm{H^\pm}\simeq0$, corresponding to the (small $\left|\varsigma_u\right|$, large $\left|\varsigma_d\right|$) region. We have then discussed their impacts on the angular observables $P_2$ and $P_5'$ in the decay $B^0\to K^{\ast 0}\mu^+\mu^-$. It is found that there is only a small impact on $P_2$ and $P_5'$ in case~A, while the case~B could obviously increase $P_5'$ to be consistent with the experimental data and reduce $P_2$ when the dimuon invariant mass squared $q^2$ is higher than the zero-crossing point.

Finally, we have explored the constraints on $\tan\beta$ and $M_{H^\pm}$ in four types of $\mathcal{Z}_2$-symmetric 2HDMs. The role of chirality-flipped operators $O_{9,10}'$ becomes much more important for large values of $\tan\beta$. Even with the current data, the types I and X and types II and Y could be clearly distinguished from each other. However, the charged-scalar effect on $P_2$ and $P_5'$ in these models is found to be small and does not help to explain the so-called $P_5'$ anomaly.

Future precise measurements of the angular observables in $b\to s\ell^+\ell^-$ decays, especially with a finer binning of $q^2$, would be very helpful to provide a more definite answer concerning the observed anomalies by the LHCb and Belle collaborations, restricting further or even deciphering the NP models.

\section*{Acknowledgements}

The work is supported by the National Natural Science Foundation of China (NSFC) under contract Nos.~11675061, 11435003, 11225523 and 11521064. Q.H. is supported by the Excellent Doctorial Dissertation Cultivation Grant from CCNU, under contract number 2013YBZD19.

\appendix

\section{Basic function}
\label{app:Basicfunction}

The basic functions $F_i(x)$ introduced in Eqs.~(\ref{eq:fi_feynman}) and (\ref{eq:fi_unitary}) are defined, respectively, as
\begin{align} \label{eq:BF1}
F_{0}(x)&=\ln x\,,\\[0.3cm] \label{eq:BF2}
F_{1}(x)&=\frac{x}{4-4x}+\frac{x\ln x}{4(x-1)^{2}}\,,\\[0.2cm] \label{eq:BF3}
F_{2}(x)&=\frac{x}{96(x-1)}-\frac{x^{2}\ln x}{96(x-1)^{2}}\,,\\[0.3cm] \label{eq:BF4}
F_{3}(x)&=\frac{x}{8}\left[\frac{x-6}{x-1}+\frac{(3x+2)\ln x}{(x-1)^{2}}\right]\,,\\[0.3cm] \label{eq:BF5}
F_{4}(x)&=-\frac{3x(x-3)}{32(x-1)}+\frac{x\left(x^{2}-8x+4\right)\ln x}{16(x-1)^{2}}\,,\\[0.3cm] \label{eq:BF6}
F_{5}(x)&=\frac{-19x^{3}+25x^{2}}{36(x-1)^{3}}+\frac{\left(5x^{2}-2x-6\right)x^{2}\ln x}{18(x-1)^{4}}\,,\\[0.3cm] \label{eq:BF7}
F_{6}(x)&=\frac{8x^{3}+5x^{2}-7x}{12(x-1)^{3}}-\frac{(3x-2)x^{2}\ln x}{2(x-1)^{4}}\,,\\[0.3cm] \label{eq:BF8}
F_{7}(x)&=\frac{x\left(53x^{2}+8x-37\right)}{108(x-1)^{4}}+\frac{x\left(-3x^{3}-9x^{2}+6x+2\right)\ln x}{18(x-1)^{5}}\,,\\[0.3cm] \label{eq:BF9}
F_{8}(x)&=\frac{x\left(18x^{4}+253x^{3}-767x^{2}+853x-417\right)}{540(x-1)^{5}} \nonumber\\[0.2cm]
&\hspace{0.6cm} -\frac{x\left(3x^{4}-6x^{3}+3x^{2}+2x-3\right)\ln x}{9(x-1)^{6}}\,.
\end{align}

\section{Wilson coefficients in A2HDM}
\label{app:WCA2HDM}

The coefficients of the different combinations of the couplings $\varsigma_u$ and $\varsigma_d$ in Eqs.~(\ref{eq:WC7inA2HDM})--(\ref{eq:WC10pinA2HDM}) are given, respectively, by
\begin{align} \label{eq:WCA2HDM1}
&C_{7,\,\mathrm{uu}}=-\frac{1}{6}F_{6}(y_{t})\,,\\[0.3cm]  \label{eq:WCA2HDM2} &C_{7,\,\mathrm{ud}}=-\frac{4}{3}F_{1}(y_{t})-\frac{80}{17}F_{2}(y_{t}) -\frac{3}{17}F_{5}(y_{t})+\frac{1}{17}F_{6}(y_{t})\,,\\[0.3cm] \label{eq:WCA2HDM3}
&C_{9,\,\mathrm{uu}}=\frac{8}{9}F_{1}(y_{t})-\frac{896}{51}F_{2}(y_{t}) -\frac{1}{17}F_{5}(y_{t})-\frac{14}{153}F_{6}(y_{t}) \, \nonumber \\[0.2cm]
& \hspace{1.4cm}-\frac{x_{t}}{2}\left(-4+\frac{1}{\sin^{2} \theta_{W}}\right)F_{1}(y_{t})\,,\\[0.3cm] \label{eq:WCA2HDM4}
&C_{10,\,\mathrm{uu}}=\frac{x_{t}}{2\sin^{2}\theta_{W}}F_{1}(y_{t})\,,\\[0.3cm] \label{eq:WCA2HDM5}
&C_{9,\,\mathrm{uu}}'=\frac{y_{t}}{x_{t}}F_{8}(y_{t})\,,\\[0.3cm] \label{eq:WCA2HDM6}
&C_{9,\,\mathrm{ud}}'=\frac{y_{t}}{x_{t}}F_{7}(y_{t})\,,\\[0.3cm] \label{eq:WCA2HDM7}
&C_{9,\,\mathrm{dd}}'=\frac{y_{t}}{x_{t}}\left[\frac{2}{9}F_{0}\left(x_{t}\right) +\frac{20}{9}F_{1}(y_{t})+\frac{928}{51}F_{2}(y_{t}) -\frac{2}{17}F_{5}(y_{t})-\frac{11}{153}F_{6}(y_{t})\right]\,,\\[0.3cm] \label{eq:WCA2HDM8}
&C_{10,\,\mathrm{uu}}'=-\frac{1}{17}\Big[80F_{2}(y_{t})+3F_{5}(y_{t}) -F_{6}(y_{t})\Big]\,,\\[0.3cm] \label{eq:WCA2HDM9}
&C_{10,\,\mathrm{ud}}'=\frac{1}{\sin^{2}\theta_{W}}\left[-\frac{1}{12}F_{1}(y_{t}) +\frac{30}{17}F_{2}(y_{t})+\frac{9}{136}F_{5}(y_{t}) -\frac{3}{136}F_{6}(y_{t})\right]\, \nonumber \\[0.2cm]
& \hspace{1.6cm} -\frac{1}{6}\left(-4+\frac{1}{\sin^{2}\theta_{W}}\right)F_{1} (y_{t})\,,\\[0.3cm] \label{eq:WCA2HDM10}
&C_{10,\,\mathrm{dd}}'=-\frac{1}{\sin^{2}\theta_{W}}\left[\frac{1}{2}F_{1}(y_{t}) +F_{2}(y_{t})\right]+\left(-4+\frac{1}{\sin^{2}\theta_{W}}\right)F_{2}(y_{t})\,,
\end{align}
and for the Wilson coefficient $C_8^\mathrm{H^\pm}$, we have~\cite{Grinstein:1990tj}
\begin{align} \label{eq:WCA2HDM11}
&C_{8,\,\mathrm{uu}}=\frac{1}{34}\Big[720F_{2}(y_{t})+27F_{5}(y_{t}) +8F_{6}(y_{t})\Big]\,,\\[0.3cm]  \label{eq:WCA2HDM12}
&C_{8,\,\mathrm{ud}}=2F_{1}(y_{t})-\frac{1}{17}\Big[240F_{2}(y_{t}) +9F_{5}(y_{t})-3F_{6}(y_{t})\Big]\,.
\end{align}

\bibliographystyle{JHEP}

\begin{thebibliography}{100}

\bibitem{Glashow:1970gm}
S.~L. Glashow, J.~Iliopoulos, and L.~Maiani, {\it {Weak Interactions with
  Lepton-Hadron Symmetry}},  {\it Phys. Rev.} {\bf D2} (1970) 1285--1292.

\bibitem{Blake:2016olu}
T.~Blake, G.~Lanfranchi, and D.~M. Straub, {\it {Rare $B$ Decays as Tests of
  the Standard Model}},  {\it Prog. Part. Nucl. Phys.} {\bf 92} (2017) 50--91,
  [\href{http://arxiv.org/abs/1606.00916}{{\tt arXiv:1606.00916}}].

\bibitem{Beneke:2001at}
M.~Beneke, T.~Feldmann and D.~Seidel, {\it {Systematic approach to exclusive $B \to  V l^+ l^-$, $V \gamma$ decays}},
  {\it Nucl. Phys.} {\bf B612} (2001) 25--58, [\href{http://arxiv.org/abs/hep-ph/0106067}{{\tt hep-ph/0106067}}].

\bibitem{Beneke:2004dp}
M.~Beneke, T.~Feldmann and D.~Seidel, {\it {Exclusive radiative and electroweak $b \to d$ and $b \to s$ penguin decays at NLO}},  {\it Eur. Phys. J.} {\bf C41} (2005) 173, [\href{http://arxiv.org/abs/hep-ph/0412400}{{\tt hep-ph/0412400}}].

\bibitem{Grinstein:2004vb}
  B.~Grinstein and D.~Pirjol, {\it {Exclusive rare $B \to K^*\ell^+\ell^-$ decays at low recoil: Controlling the long-distance effects}},
  {\it Phys. Rev.} {\bf D70} (2004) 114005, [\href{http://arxiv.org/abs/hep-ph/0404250}{{\tt hep-ph/0404250}}].

\bibitem{Altmannshofer:2008dz}
W.~Altmannshofer, P.~Ball, A.~Bharucha, A.~J. Buras, D.~M. Straub, and M.~Wick,
  {\it {Symmetries and Asymmetries of $B \to K^{*} \mu^{+} \mu^{-}$ Decays in
  the Standard Model and Beyond}},  {\it JHEP} {\bf 01} (2009) 019,
  [\href{http://arxiv.org/abs/0811.1214}{{\tt arXiv:0811.1214}}].

\bibitem{Beylich:2011aq}
M.~Beylich, G.~Buchalla and T.~Feldmann, {\it {Theory of $B \to K^{(*)}\ell^+
  \ell^-$ decays at high $q^2$: OPE and quark-hadron duality}},  {\it Eur. Phys. J.} {\bf C71} (2011) 1635, [\href{http://arxiv.org/abs/1101.5118}{{\tt arXiv:1101.5118}}].

\bibitem{DescotesGenon:2012zf}
S.~Descotes-Genon, J.~Matias, M.~Ramon, and J.~Virto, {\it {Implications from
  clean observables for the binned analysis of $B \to K^\ast\mu^+\mu^-$ at
  large recoil}},  {\it JHEP} {\bf 01} (2013) 048,
  [\href{http://arxiv.org/abs/1207.2753}{{\tt arXiv:1207.2753}}].

\bibitem{Descotes-Genon:2013vna}
S.~Descotes-Genon, T.~Hurth, J.~Matias, and J.~Virto, {\it {Optimizing the
  basis of $B\to K^\ast\ell^+\ell^-$ observables in the full kinematic range}},
   {\it JHEP} {\bf 05} (2013) 137, [\href{http://arxiv.org/abs/1303.5794}{{\tt
  arXiv:1303.5794}}].

\bibitem{Gratrex:2015hna}
J.~Gratrex, M.~Hopfer, and R.~Zwicky, {\it {Generalised helicity formalism,
  higher moments and the $B \to K_{J_K}(\to K \pi) \bar{\ell}_1 \ell_2$ angular
  distributions}},  {\it Phys. Rev.} {\bf D93} (2016), no.~5 054008,
  [\href{http://arxiv.org/abs/1506.03970}{{\tt arXiv:1506.03970}}].

\bibitem{Aaij:2013qta}
{\bf LHCb} Collaboration, R.~Aaij et~al., {\it {Measurement of
  Form-Factor-Independent Observables in the Decay $B^{0} \to K^{*0} \mu^+
  \mu^-$}},  {\it Phys. Rev. Lett.} {\bf 111} (2013) 191801,
  [\href{http://arxiv.org/abs/1308.1707}{{\tt arXiv:1308.1707}}].

\bibitem{Descotes-Genon:2014uoa}
S.~Descotes-Genon, L.~Hofer, J.~Matias, and J.~Virto, {\it {On the impact of
  power corrections in the prediction of $B \to K^*\mu^+\mu^-$ observables}},
  {\it JHEP} {\bf 12} (2014) 125, [\href{http://arxiv.org/abs/1407.8526}{{\tt
  arXiv:1407.8526}}].

\bibitem{Straub:2015ica}
A.~Bharucha, D.~M. Straub, and R.~Zwicky, {\it {$B\to V\ell^+\ell^-$ in the
  Standard Model from light-cone sum rules}},  {\it JHEP} {\bf 08} (2016) 098,
  [\href{http://arxiv.org/abs/1503.05534}{{\tt arXiv:1503.05534}}].

\bibitem{Jager:2012uw}
S.~J\"{a}ger and J.~Martin~Camalich, {\it {On $B \to V \ell \ell$ at small dilepton
  invariant mass, power corrections, and new physics}},  {\it JHEP} {\bf 05}
  (2013) 043, [\href{http://arxiv.org/abs/1212.2263}{{\tt arXiv:1212.2263}}].

\bibitem{Jager:2014rwa}
S.~J\"{a}ger and J.~Martin~Camalich, {\it {Reassessing the discovery potential of
  the $B \to K^{*} \ell^+\ell^-$ decays in the large-recoil region: SM
  challenges and BSM opportunities}},  {\it Phys. Rev.} {\bf D93} (2016), no.~1
  014028, [\href{http://arxiv.org/abs/1412.3183}{{\tt arXiv:1412.3183}}].

\bibitem{Aaij:2015oid}
{\bf LHCb} Collaboration, R.~Aaij et~al., {\it {Angular analysis of the $B^{0}
  \to K^{*0} \mu^{+} \mu^{-}$ decay using 3 fb$^{-1}$ of integrated
  luminosity}},  {\it JHEP} {\bf 02} (2016) 104,
  [\href{http://arxiv.org/abs/1512.04442}{{\tt arXiv:1512.04442}}].

\bibitem{Abdesselam:2016llu}
{\bf Belle} Collaboration, A.~Abdesselam et~al., {\it {Angular analysis of $B^0
  \to K^\ast(892)^0 \ell^+ \ell^-$}},  in {\it {Proceedings, LHCSki 2016 - A
  First Discussion of 13 TeV Results: Obergurgl, Austria, April 10-15, 2016}},
  2016.
\newblock \href{http://arxiv.org/abs/1604.04042}{{\tt arXiv:1604.04042}}.

\bibitem{Altmannshofer:2014rta}
W.~Altmannshofer and D.~M. Straub, {\it {New physics in $b\rightarrow s$
  transitions after LHC run 1}},  {\it Eur. Phys. J.} {\bf C75} (2015), no.~8
  382, [\href{http://arxiv.org/abs/1411.3161}{{\tt arXiv:1411.3161}}].

\bibitem{Descotes-Genon:2015uva}
S.~Descotes-Genon, L.~Hofer, J.~Matias, and J.~Virto, {\it {Global analysis of
  $b\to s\ell\ell$ anomalies}},  {\it JHEP} {\bf 06} (2016) 092,
  [\href{http://arxiv.org/abs/1510.04239}{{\tt arXiv:1510.04239}}].

\bibitem{Hurth:2016fbr}
T.~Hurth, F.~Mahmoudi, and S.~Neshatpour, {\it {On the anomalies in the latest
  LHCb data}},  {\it Nucl. Phys.} {\bf B909} (2016) 737--777,
  [\href{http://arxiv.org/abs/1603.00865}{{\tt arXiv:1603.00865}}].

\bibitem{Hurth:2013ssa}
T.~Hurth and F.~Mahmoudi, {\it {On the LHCb anomaly in B $\to
  K^*\ell^+\ell^-$}},  {\it JHEP} {\bf 04} (2014) 097,
  [\href{http://arxiv.org/abs/1312.5267}{{\tt arXiv:1312.5267}}].

\bibitem{Descotes-Genon:2013wba}
S.~Descotes-Genon, J.~Matias, and J.~Virto, {\it {Understanding the $B\to
  K^*\mu^+\mu^-$ Anomaly}},  {\it Phys. Rev.} {\bf D88} (2013) 074002,
  [\href{http://arxiv.org/abs/1307.5683}{{\tt arXiv:1307.5683}}].

\bibitem{Altmannshofer:2013foa}
W.~Altmannshofer and D.~M. Straub, {\it {New Physics in $B \to K^*\mu\mu$?}},
  {\it Eur. Phys. J.} {\bf C73} (2013) 2646,
  [\href{http://arxiv.org/abs/1308.1501}{{\tt arXiv:1308.1501}}].

\bibitem{Beaujean:2013soa}
F.~Beaujean, C.~Bobeth, and D.~van Dyk, {\it {Comprehensive Bayesian analysis
  of rare (semi)leptonic and radiative $B$ decays}},  {\it Eur. Phys. J.} {\bf
  C74} (2014) 2897, [\href{http://arxiv.org/abs/1310.2478}{{\tt
  arXiv:1310.2478}}]. [Erratum: Eur. Phys. J.C74,3179(2014)].

\bibitem{Horgan:2013pva}
R.~R. Horgan, Z.~Liu, S.~Meinel, and M.~Wingate, {\it {Calculation of $B^0 \to
  K^{*0} \mu^+ \mu^-$ and $B_s^0 \to \phi \mu^+ \mu^-$ observables using form
  factors from lattice QCD}},  {\it Phys. Rev. Lett.} {\bf 112} (2014) 212003,
  [\href{http://arxiv.org/abs/1310.3887}{{\tt arXiv:1310.3887}}].

\bibitem{Hurth:2014vma}
T.~Hurth, F.~Mahmoudi, and S.~Neshatpour, {\it {Global fits to $b \to
  s\ell\ell$ data and signs for lepton non-universality}},  {\it JHEP} {\bf 12}
  (2014) 053, [\href{http://arxiv.org/abs/1410.4545}{{\tt arXiv:1410.4545}}].

\bibitem{Du:2015tda}
D.~Du, A.~X. El-Khadra, S.~Gottlieb, A.~S. Kronfeld, J.~Laiho, E.~Lunghi, R.~S.
  Van~de Water, and R.~Zhou, {\it {Phenomenology of semileptonic B-meson decays
  with form factors from lattice QCD}},  {\it Phys. Rev.} {\bf D93} (2016),
  no.~3 034005, [\href{http://arxiv.org/abs/1510.02349}{{\tt
  arXiv:1510.02349}}].

\bibitem{Ciuchini:2015qxb}
M.~Ciuchini, M.~Fedele, E.~Franco, S.~Mishima, A.~Paul, L.~Silvestrini, and
  M.~Valli, {\it {$B\to K^* \ell^+ \ell^-$ decays at large recoil in the
  Standard Model: a theoretical reappraisal}},  {\it JHEP} {\bf 06} (2016) 116,
  [\href{http://arxiv.org/abs/1512.07157}{{\tt arXiv:1512.07157}}].

\bibitem{Meinel:2016grj}
S.~Meinel and D.~van Dyk, {\it {Using $\Lambda_b\to \Lambda\mu^+\mu^-$ data
  within a Bayesian analysis of $|\Delta B| = |\Delta S| = 1$ decays}},  {\it
  Phys. Rev.} {\bf D94} (2016), no.~1 013007,
  [\href{http://arxiv.org/abs/1603.02974}{{\tt arXiv:1603.02974}}].

\bibitem{Khodjamirian:2010vf}
A.~Khodjamirian, T.~Mannel, A.~A. Pivovarov, and Y.~M. Wang, {\it {Charm-loop
  effect in $B \to K^{(*)} \ell^{+} \ell^{-}$ and $B\to K^*\gamma$}},  {\it
  JHEP} {\bf 09} (2010) 089, [\href{http://arxiv.org/abs/1006.4945}{{\tt
  arXiv:1006.4945}}].

\bibitem{Brass:2016efg}
S.~Bra{\ss}, G.~Hiller, and I.~Nisandzic, {\it {Zooming in on $B\rightarrow
  K^*\ell \ell $ decays at low recoil}},  {\it Eur. Phys. J.} {\bf C77} (2017),
  no.~1 16, [\href{http://arxiv.org/abs/1606.00775}{{\tt arXiv:1606.00775}}].

\bibitem{Capdevila:2016ivx}
B.~Capdevila, S.~Descotes-Genon, J.~Matias, and J.~Virto, {\it {Assessing
  lepton-flavour non-universality from $B\to K^*\ell\ell$ angular analyses}},
  {\it JHEP} {\bf 10} (2016) 075, [\href{http://arxiv.org/abs/1605.03156}{{\tt
  arXiv:1605.03156}}].

\bibitem{Karan:2016wvu}
A.~Karan, R.~Mandal, A.~K. Nayak, R.~Sinha, and T.~E. Browder, {\it {Signal of
  right-handed currents using $B\to K^*\ell^+\ell^-$ observables at the
  kinematic endpoint}},  \href{http://arxiv.org/abs/1603.04355}{{\tt
  arXiv:1603.04355}}.

\bibitem{Ahmed:2016jgv}
I.~Ahmed, M.~J. Aslam, and M.~A. Paracha, {\it {Asymmetries in $B \to K^\ast
  \ell^+ \ell^-$ Decays and Two Higgs Doublet Model}},
  \href{http://arxiv.org/abs/1602.02400}{{\tt arXiv:1602.02400}}.

\bibitem{Chiang:2016qov}
C.-W. Chiang, X.-G. He, and G.~Valencia, {\it {$Z^\prime$ model for $b \to s
  \ell \bar\ell$ flavor anomalies}},  {\it Phys. Rev.} {\bf D93} (2016), no.~7
  074003, [\href{http://arxiv.org/abs/1601.07328}{{\tt arXiv:1601.07328}}].

\bibitem{Celis:2015eqs}
A.~Celis, W.-Z. Feng, and D.~L{\"u}st, {\it {Stringy explanation of $b\to s
  \ell^{+}\ell^{-}$ anomalies}},  {\it JHEP} {\bf 02} (2016) 007,
  [\href{http://arxiv.org/abs/1512.02218}{{\tt arXiv:1512.02218}}].

\bibitem{Boucenna:2016wpr}
S.~M. Boucenna, A.~Celis, J.~Fuentes-Martin, A.~Vicente, and J.~Virto, {\it
  {Non-abelian gauge extensions for B-decay anomalies}},  {\it Phys. Lett.}
  {\bf B760} (2016) 214--219, [\href{http://arxiv.org/abs/1604.03088}{{\tt
  arXiv:1604.03088}}].

\bibitem{Crivellin:2016ejn}
A.~Crivellin, J.~Fuentes-Martin, A.~Greljo, and G.~Isidori, {\it {Lepton Flavor
  Non-Universality in B decays from Dynamical Yukawas}},  {\it Phys. Lett.}
  {\bf B766} (2017) 77--85, [\href{http://arxiv.org/abs/1611.02703}{{\tt
  arXiv:1611.02703}}].

\bibitem{Barbieri:2016las}
R.~Barbieri, C.~W. Murphy, and F.~Senia, {\it {B-decay Anomalies in a Composite
  Leptoquark Model}},  {\it Eur. Phys. J.} {\bf C77} (2017), no.~1 8,
  [\href{http://arxiv.org/abs/1611.04930}{{\tt arXiv:1611.04930}}].

\bibitem{Mahmoudi:2016mgr}
F.~Mahmoudi, T.~Hurth, and S.~Neshatpour, {\it {Present Status of $b \to s \ell^+ \ell^-$ Anomalies}},  2016.
\newblock \href{http://arxiv.org/abs/1611.05060}{{\tt arXiv:1611.05060}}.

\bibitem{Crivellin:2015mga}
A.~Crivellin, G.~D'Ambrosio, and J.~Heeck, {\it {Explaining
  $h\to\mu^\pm\tau^\mp$, $B\to K^* \mu^+\mu^-$ and $B\to K \mu^+\mu^-/B\to K
  e^+e^-$ in a two-Higgs-doublet model with gauged $L_\mu-L_\tau$}},  {\it
  Phys. Rev. Lett.} {\bf 114} (2015) 151801,
  [\href{http://arxiv.org/abs/1501.00993}{{\tt arXiv:1501.00993}}].

\bibitem{Crivellin:2015lwa}
A.~Crivellin, G.~D'Ambrosio, and J.~Heeck, {\it {Addressing the LHC flavor
  anomalies with horizontal gauge symmetries}},  {\it Phys. Rev.} {\bf D91}
  (2015), no.~7 075006, [\href{http://arxiv.org/abs/1503.03477}{{\tt
  arXiv:1503.03477}}].

\bibitem{Calibbi:2015kma}
L.~Calibbi, A.~Crivellin, and T.~Ota, {\it {Effective Field Theory Approach to
  $b\to s\ell\ell^{(\prime)}$, $B\to K^{(\ast)}\nu\bar\nu$ and $B\to
  D^{(\ast)}\tau\nu$ with Third Generation Couplings}},  {\it Phys. Rev. Lett.}
  {\bf 115} (2015) 181801, [\href{http://arxiv.org/abs/1506.02661}{{\tt
  arXiv:1506.02661}}].

\bibitem{Arnan:2016cpy}
P.~Arnan, L.~Hofer, F.~Mescia, and A.~Crivellin, {\it {Loop effects of heavy
  new scalars and fermions in $b\to s\mu^+\mu^-$}},
  \href{http://arxiv.org/abs/1608.07832}{{\tt arXiv:1608.07832}}.

\bibitem{Lee:1973iz}
T.~D. Lee, {\it {A Theory of Spontaneous T Violation}},  {\it Phys. Rev.} {\bf
  D8} (1973) 1226--1239.

\bibitem{Branco:2011iw}
G.~C. Branco, P.~M. Ferreira, L.~Lavoura, M.~N. Rebelo, M.~Sher, and J.~P.
  Silva, {\it {Theory and phenomenology of two-Higgs-doublet models}},  {\it
  Phys. Rept.} {\bf 516} (2012) 1--102,
  [\href{http://arxiv.org/abs/1106.0034}{{\tt arXiv:1106.0034}}].

\bibitem{Aad:2012tfa}
{\bf ATLAS} Collaboration, G.~Aad et~al., {\it {Observation of a new particle
  in the search for the Standard Model Higgs boson with the ATLAS detector at
  the LHC}},  {\it Phys. Lett.} {\bf B716} (2012) 1--29,
  [\href{http://arxiv.org/abs/1207.7214}{{\tt arXiv:1207.7214}}].

\bibitem{Chatrchyan:2012xdj}
{\bf CMS} Collaboration, S.~Chatrchyan et~al., {\it {Observation of a new boson
  at a mass of 125 GeV with the CMS experiment at the LHC}},  {\it Phys. Lett.}
  {\bf B716} (2012) 30--61, [\href{http://arxiv.org/abs/1207.7235}{{\tt
  arXiv:1207.7235}}].

\bibitem{Haber:1984rc}
H.~E. Haber and G.~L. Kane, {\it {The Search for Supersymmetry: Probing Physics
  Beyond the Standard Model}},  {\it Phys. Rept.} {\bf 117} (1985) 75--263.

\bibitem{Kim:1986ax}
J.~E. Kim, {\it {Light Pseudoscalars, Particle Physics and Cosmology}},  {\it
  Phys. Rept.} {\bf 150} (1987) 1--177.

\bibitem{Trodden:1998qg}
M.~Trodden, {\it {Electroweak baryogenesis: A Brief review}},  in {\it
  {Proceedings, 33rd Rencontres de Moriond 98 electrowek interactions and
  unified theories: Les racs, France, Mar 14-21, 1998}}, pp.~471--480, 1998.
\newblock \href{http://arxiv.org/abs/hep-ph/9805252}{{\tt hep-ph/9805252}}.

\bibitem{Gunion:1989we}
J.~F. Gunion, H.~E. Haber, G.~L. Kane, and S.~Dawson, {\it {The Higgs Hunter's
  Guide}},  {\it Front. Phys.} {\bf 80} (2000) 1--404.

\bibitem{Glashow:1976nt}
S.~L. Glashow and S.~Weinberg, {\it {Natural Conservation Laws for Neutral
  Currents}},  {\it Phys. Rev.} {\bf D15} (1977) 1958.

\bibitem{Pich:2009sp}
A.~Pich and P.~Tuz\'on, {\it {Yukawa Alignment in the Two-Higgs-Doublet
  Model}},  {\it Phys. Rev.} {\bf D80} (2009) 091702,
  [\href{http://arxiv.org/abs/0908.1554}{{\tt arXiv:0908.1554}}].

\bibitem{Cabibbo:1963yz}
N.~Cabibbo, {\it {Unitary Symmetry and Leptonic Decays}},  {\it Phys. Rev.
  Lett.} {\bf 10} (1963) 531--533.

\bibitem{Kobayashi:1973fv}
M.~Kobayashi and T.~Maskawa, {\it {CP Violation in the Renormalizable Theory of
  Weak Interaction}},  {\it Prog. Theor. Phys.} {\bf 49} (1973) 652--657.

\bibitem{Altmannshofer:2012ar}
W.~Altmannshofer, S.~Gori, and G.~D. Kribs, {\it {A Minimal Flavor Violating
  2HDM at the LHC}},  {\it Phys. Rev.} {\bf D86} (2012) 115009,
  [\href{http://arxiv.org/abs/1210.2465}{{\tt arXiv:1210.2465}}].

\bibitem{Bai:2012ex}
Y.~Bai, V.~Barger, L.~L. Everett, and G.~Shaughnessy, {\it {General two Higgs
  doublet model (2HDM-G) and Large Hadron Collider data}},  {\it Phys. Rev.}
  {\bf D87} (2013) 115013, [\href{http://arxiv.org/abs/1210.4922}{{\tt
  arXiv:1210.4922}}].

\bibitem{Barger:2013ofa}
V.~Barger, L.~L. Everett, H.~E. Logan, and G.~Shaughnessy, {\it {Scrutinizing
  the 125 GeV Higgs boson in two Higgs doublet models at the LHC, ILC, and Muon
  Collider}},  {\it Phys. Rev.} {\bf D88} (2013), no.~11 115003,
  [\href{http://arxiv.org/abs/1308.0052}{{\tt arXiv:1308.0052}}].

\bibitem{Lopez-Val:2013yba}
D.~L\'opez-Val, T.~Plehn, and M.~Rauch, {\it {Measuring extended Higgs sectors
  as a consistent free couplings model}},  {\it JHEP} {\bf 10} (2013) 134,
  [\href{http://arxiv.org/abs/1308.1979}{{\tt arXiv:1308.1979}}].

\bibitem{Wang:2013sha}
L.~Wang and X.-F. Han, {\it {Status of the aligned two-Higgs-doublet model
  confronted with the Higgs data}},  {\it JHEP} {\bf 04} (2014) 128,
  [\href{http://arxiv.org/abs/1312.4759}{{\tt arXiv:1312.4759}}].

\bibitem{Celis:2013rcs}
A.~Celis, V.~Ilisie, and A.~Pich, {\it {LHC constraints on two-Higgs doublet
  models}},  {\it JHEP} {\bf 07} (2013) 053,
  [\href{http://arxiv.org/abs/1302.4022}{{\tt arXiv:1302.4022}}].

\bibitem{Celis:2013ixa}
A.~Celis, V.~Ilisie, and A.~Pich, {\it {Towards a general analysis of LHC data
  within two-Higgs-doublet models}},  {\it JHEP} {\bf 12} (2013) 095,
  [\href{http://arxiv.org/abs/1310.7941}{{\tt arXiv:1310.7941}}].

\bibitem{Jung:2010ik}
M.~Jung, A.~Pich, and P.~Tuz\'on, {\it {Charged-Higgs phenomenology in the
  Aligned two-Higgs-doublet model}},  {\it JHEP} {\bf 11} (2010) 003,
  [\href{http://arxiv.org/abs/1006.0470}{{\tt arXiv:1006.0470}}].

\bibitem{Jung:2010ab}
M.~Jung, A.~Pich, and P.~Tuz\'on, {\it {The $\bar B\to X_s\gamma$ Rate and CP
  Asymmetry within the Aligned Two-Higgs-Doublet Model}},  {\it Phys. Rev.}
  {\bf D83} (2011) 074011, [\href{http://arxiv.org/abs/1011.5154}{{\tt
  arXiv:1011.5154}}].

\bibitem{Jung:2012vu}
M.~Jung, X.-Q. Li, and A.~Pich, {\it {Exclusive radiative B-meson decays within
  the aligned two-Higgs-doublet model}},  {\it JHEP} {\bf 10} (2012) 063,
  [\href{http://arxiv.org/abs/1208.1251}{{\tt arXiv:1208.1251}}].

\bibitem{Celis:2012dk}
A.~Celis, M.~Jung, X.-Q. Li, and A.~Pich, {\it {Sensitivity to charged scalars
  in $B\to D^{(*)}\tau\nu_\tau$ and $B\to\tau\nu_\tau$ decays}},  {\it JHEP}
  {\bf 01} (2013) 054, [\href{http://arxiv.org/abs/1210.8443}{{\tt
  arXiv:1210.8443}}].

\bibitem{Duarte:2013zfa}
L.~Duarte, G.~A. Gonz\'alez-Sprinberg, and J.~Vidal, {\it {Top quark anomalous
  tensor couplings in the two-Higgs-doublet models}},  {\it JHEP} {\bf 11}
  (2013) 114, [\href{http://arxiv.org/abs/1308.3652}{{\tt arXiv:1308.3652}}].

\bibitem{Jung:2013hka}
M.~Jung and A.~Pich, {\it {Electric Dipole Moments in Two-Higgs-Doublet
  Models}},  {\it JHEP} {\bf 04} (2014) 076,
  [\href{http://arxiv.org/abs/1308.6283}{{\tt arXiv:1308.6283}}].

\bibitem{Li:2014fea}
X.-Q. Li, J.~Lu, and A.~Pich, {\it {$B_{s,d}^0 \to \ell^+\ell^-$ Decays in the
  Aligned Two-Higgs-Doublet Model}},  {\it JHEP} {\bf 06} (2014) 022,
  [\href{http://arxiv.org/abs/1404.5865}{{\tt arXiv:1404.5865}}].

\bibitem{Ilisie:2015tra}
V.~Ilisie, {\it {New Barr-Zee contributions to $(g-2)_\mu$ in two-Higgs-doublet
  models}},  {\it JHEP} {\bf 04} (2015) 077,
  [\href{http://arxiv.org/abs/1502.04199}{{\tt arXiv:1502.04199}}].

\bibitem{Abbas:2015cua}
G.~Abbas, A.~Celis, X.-Q. Li, J.~Lu, and A.~Pich, {\it {Flavour-changing top
  decays in the aligned two-Higgs-doublet model}},  {\it JHEP} {\bf 06} (2015)
  005, [\href{http://arxiv.org/abs/1503.06423}{{\tt arXiv:1503.06423}}].

\bibitem{Han:2015yys}
T.~Han, S.~K. Kang, and J.~Sayre, {\it {Muon $g-2$ in the aligned two Higgs
  doublet model}},  {\it JHEP} {\bf 02} (2016) 097,
  [\href{http://arxiv.org/abs/1511.05162}{{\tt arXiv:1511.05162}}].

\bibitem{Wang:2016rvz}
L.~Wang, S.~Yang, and X.-F. Han, {\it {$h\to\mu\tau$ and muon g-2 in the
  alignment limit of two-Higgs-doublet model}},
  \href{http://arxiv.org/abs/1606.04408}{{\tt arXiv:1606.04408}}.

\bibitem{Davidson:2005cw}
S.~Davidson and H.~E. Haber, {\it {Basis-independent methods for the
  two-Higgs-doublet model}},  {\it Phys. Rev.} {\bf D72} (2005) 035004,
  [\href{http://arxiv.org/abs/hep-ph/0504050}{{\tt hep-ph/0504050}}]. [Erratum:
  Phys. Rev.D72,099902(2005)].

\bibitem{Haber:2006ue}
H.~E. Haber and D.~O'Neil, {\it {Basis-independent methods for the
  two-Higgs-doublet model. II. The Significance of tan$\beta$}},  {\it Phys.
  Rev.} {\bf D74} (2006) 015018,
  [\href{http://arxiv.org/abs/hep-ph/0602242}{{\tt hep-ph/0602242}}]. [Erratum:
  Phys. Rev.D74,no.5,059905(2006)].

\bibitem{Haber:2010bw}
H.~E. Haber and D.~O'Neil, {\it {Basis-independent methods for the
  two-Higgs-doublet model III: The CP-conserving limit, custodial symmetry, and
  the oblique parameters S, T, U}},  {\it Phys. Rev.} {\bf D83} (2011) 055017,
  [\href{http://arxiv.org/abs/1011.6188}{{\tt arXiv:1011.6188}}].

\bibitem{Buchalla:1995vs}
G.~Buchalla, A.~J. Buras, and M.~E. Lautenbacher, {\it {Weak decays beyond
  leading logarithms}},  {\it Rev. Mod. Phys.} {\bf 68} (1996) 1125--1144,
  [\href{http://arxiv.org/abs/hep-ph/9512380}{{\tt hep-ph/9512380}}].

\bibitem{Inami:1980fz}
T.~Inami and C.~S. Lim, {\it {Effects of Superheavy Quarks and Leptons in
  Low-Energy Weak Processes $K_L\to \mu{\bar \mu}$, $K^+\to \pi^+\nu{\bar \nu}$
  and $K^0\leftrightarrow{\bar K}^0$}},  {\it Prog. Theor. Phys.} {\bf 65}
  (1981) 297. [Erratum: Prog. Theor. Phys.65,1772(1981)].

\bibitem{Misiak:1992bc}
M.~Misiak, {\it {The $b \to se^+ e^-$ and $b \to s\gamma$ decays with
  next-to-leading logarithmic QCD corrections}},  {\it Nucl. Phys.} {\bf B393}
  (1993) 23--45. [Erratum: Nucl. Phys.B439,461(1995)].

\bibitem{Deshpande:1981zq}
N.~G. Deshpande and G.~Eilam, {\it {Flavor-changing electromagnetic
  transitions}},  {\it Phys. Rev.} {\bf D26} (1982) 2463.

\bibitem{Deshpande:1982mi}
N.~G. Deshpande and M.~Nazerimonfared, {\it {Flavor Changing Electromagnetic
  Vertex in a Nonlinear $R_\xi$ Gauge}},  {\it Nucl. Phys.} {\bf B213} (1983)
  390--408.

\bibitem{Chia:1983hd}
S.-P. Chia, {\it {An Exact Calculation of $\bar{d} s g$ Vertex}},  {\it Phys.
  Lett.} {\bf B130} (1983) 315--320.

\bibitem{Chia:1985dx}
S.-P. Chia and G.~Rajagopal, {\it {An Exact Calculation of the Flavor Changing
  Quark - Photon Vertex}},  {\it Phys. Lett.} {\bf B156} (1985) 405--410.

\bibitem{Chia:1989gh}
S.-P. Chia, {\it {Radiative Decay of the Bottom Quark and the $W W \gamma$
  Coupling}},  {\it Phys. Lett.} {\bf B240} (1990) 465--470.

\bibitem{Wu:2006sp}
L.-s. Wu and Z.-j. Xiao, {\it {Exact Calculations of Vertex $\bar s\gamma b$
  and $\bar s Z b$ in the Unitary Gauge}},  {\it Commun. Theor. Phys.} {\bf 48}
  (2007) 502--508, [\href{http://arxiv.org/abs/hep-ph/0612326}{{\tt
  hep-ph/0612326}}].

\bibitem{He:2009rz}
X.-G. He, J.~Tandean, and G.~Valencia, {\it {Penguin and Box Diagrams in
  Unitary Gauge}},  {\it Eur. Phys. J.} {\bf C64} (2009) 681--687,
  [\href{http://arxiv.org/abs/0909.3638}{{\tt arXiv:0909.3638}}].

\bibitem{Buras:2011we}
A.~J. Buras, {\it {Climbing NLO and NNLO Summits of Weak Decays}},
  \href{http://arxiv.org/abs/1102.5650}{{\tt arXiv:1102.5650}}.

\bibitem{Grinstein:1990tj}
B.~Grinstein, R.~P. Springer, and M.~B. Wise, {\it {Strong Interaction Effects
  in Weak Radiative $\bar{B}$ Meson Decay}},  {\it Nucl. Phys.} {\bf B339}
  (1990) 269--309.

\bibitem{Bertolini:1990if}
S.~Bertolini, F.~Borzumati, A.~Masiero, and G.~Ridolfi, {\it {Effects of
  supergravity induced electroweak breaking on rare $B$ decays and mixings}},
  {\it Nucl. Phys.} {\bf B353} (1991) 591--649.

\bibitem{Cho:1996we}
P.~L. Cho, M.~Misiak, and D.~Wyler, {\it {$K_L\to\pi^0 e^+ e^-$ and $B\to
  X_sl^+l^-$ decay in the MSSM}},  {\it Phys. Rev.} {\bf D54} (1996)
  3329--3344, [\href{http://arxiv.org/abs/hep-ph/9601360}{{\tt
  hep-ph/9601360}}].

\bibitem{Chankowski:2000ng}
P.~H. Chankowski and L.~Slawianowska, {\it {$B^0_{d,s}\to\mu^-\mu^+$ decay in
  the MSSM}},  {\it Phys. Rev.} {\bf D63} (2001) 054012,
  [\href{http://arxiv.org/abs/hep-ph/0008046}{{\tt hep-ph/0008046}}].

\bibitem{Ciuchini:1997xe}
M.~Ciuchini, G.~Degrassi, P.~Gambino, and G.~F. Giudice, {\it {Next-to-leading
  QCD corrections to $B \to X_s \gamma$: Standard model and two Higgs doublet
  model}},  {\it Nucl. Phys.} {\bf B527} (1998) 21--43,
  [\href{http://arxiv.org/abs/hep-ph/9710335}{{\tt hep-ph/9710335}}].

\bibitem{Borzumati:1998tg}
F.~Borzumati and C.~Greub, {\it {Two Higgs doublet model predictions for $\bar
  B\to X_s\gamma$ in NLO QCD}},  {\it Phys. Rev.} {\bf D58} (1998) 074004,
  [\href{http://arxiv.org/abs/hep-ph/9802391}{{\tt hep-ph/9802391}}].
  [Addendum: Phys. Rev.{\bf D59} (1999) 057501].

\bibitem{Bobeth:1999ww}
C.~Bobeth, M.~Misiak, and J.~Urban, {\it {Matching conditions for $b \to s
  \gamma$ and $b \to s gluon$ in extensions of the standard model}},  {\it
  Nucl. Phys.} {\bf B567} (2000) 153--185,
  [\href{http://arxiv.org/abs/hep-ph/9904413}{{\tt hep-ph/9904413}}].

\bibitem{Bobeth:2001jm}
C.~Bobeth, A.~J. Buras, F.~Kr{\"u}ger, and J.~Urban, {\it {QCD corrections to
  $\bar{B} \to X_{d,s} \nu \bar{\nu}$, $\bar{B}_{d,s} \to \ell^{+} \ell^{-}$,
  $K \to \pi \nu \bar{\nu}$ and $K_{L} \to \mu^{+} \mu^{-}$ in the MSSM}},
  {\it Nucl. Phys.} {\bf B630} (2002) 87--131,
  [\href{http://arxiv.org/abs/hep-ph/0112305}{{\tt hep-ph/0112305}}].

\bibitem{Schilling:2004gk}
S.~Schilling, C.~Greub, N.~Salzmann, and B.~T{\"o}edtli, {\it {QCD corrections
  to the Wilson coefficients $C_9$ and $C_{10}$ in two-Higgs doublet models}},
  {\it Phys. Lett.} {\bf B616} (2005) 93--100,
  [\href{http://arxiv.org/abs/hep-ph/0407323}{{\tt hep-ph/0407323}}].

\bibitem{Kruger:1999xa}
F.~Kr{\"u}ger, L.~M. Sehgal, N.~Sinha, and R.~Sinha, {\it {Angular distribution
  and CP asymmetries in the decays $\bar B\to K^-\pi^+e^-e^+$ and $\bar
  B\to\pi^-\pi^+e^-e^+$}},  {\it Phys. Rev.} {\bf D61} (2000) 114028,
  [\href{http://arxiv.org/abs/hep-ph/9907386}{{\tt hep-ph/9907386}}]. [Erratum:
  Phys. Rev.D63,019901(2001)].

\bibitem{Becirevic:2011bp}
D.~Be\v{c}irevi\'c and E.~Schneider, {\it {On transverse asymmetries in $B\to
  K^\ast\ell^+\ell^-$}},  {\it Nucl. Phys.} {\bf B854} (2012) 321--339,
  [\href{http://arxiv.org/abs/1106.3283}{{\tt arXiv:1106.3283}}].

\bibitem{Matias:2012xw}
J.~Matias, F.~Mescia, M.~Ramon, and J.~Virto, {\it {Complete Anatomy of
  $\bar{B}_d \to \bar{K}^{* 0} (\to K \pi)\ell^+\ell^-$ and its angular
  distribution}},  {\it JHEP} {\bf 04} (2012) 104,
  [\href{http://arxiv.org/abs/1202.4266}{{\tt arXiv:1202.4266}}].

\bibitem{Searches:2001ac}
{\bf OPAL, DELPHI, L3, ALEPH, LEP Higgs Working Group for Higgs boson searches}
  Collaboration, {\it {Search for charged Higgs bosons: Preliminary combined
  results using LEP data collected at energies up to 209-GeV}},  in {\it
  {Lepton and photon interactions at high energies. Proceedings, 20th
  International Symposium, LP 2001, Rome, Italy, July 23-28, 2001}}, 2001.
\newblock \href{http://arxiv.org/abs/hep-ex/0107031}{{\tt hep-ex/0107031}}.

\bibitem{Barroso:2013awa}
A.~Barroso, P.~M. Ferreira, I.~P. Ivanov, and R.~Santos, {\it {Metastability
  bounds on the two Higgs doublet model}},  {\it JHEP} {\bf 06} (2013) 045,
  [\href{http://arxiv.org/abs/1303.5098}{{\tt arXiv:1303.5098}}].

\bibitem{Dev:2014yca}
P.~S.~Bhupal Dev and A.~Pilaftsis, {\it {Maximally Symmetric Two Higgs Doublet
  Model with Natural Standard Model Alignment}},  {\it JHEP} {\bf 12} (2014) 024,
  [\href{http://arxiv.org/abs/1408.3405}{{\tt arXiv:1408.3405}}].
  [Erratum: JHEP1511,147(2015)].

\bibitem{Das:2015qva}
D.~Das, {\it {New limits on tan $\beta$ for 2HDMs with Z$_2$ symmetry}},  {\it
  Int. J. Mod. Phys.} {\bf A30} (2015), no.~26 1550158,
  [\href{http://arxiv.org/abs/1501.02610}{{\tt arXiv:1501.02610}}].

\bibitem{Chakraborty:2015raa}
I.~Chakraborty and A.~Kundu, {\it {Scalar potential of two-Higgs doublet
  models}},  {\it Phys. Rev.} {\bf D92} (2015), no.~9 095023,
  [\href{http://arxiv.org/abs/1508.00702}{{\tt arXiv:1508.00702}}].

\bibitem{Olive:2016xmw}
C.~Patrignani, {\it {Review of Particle Physics}},  {\it Chin. Phys.} {\bf C40}
  (2016), no.~10 100001.

\bibitem{Chen:2001fja}
{\bf CLEO} Collaboration, S.~Chen et~al., {\it {Branching fraction and photon
  energy spectrum for $b \to s \gamma$}},  {\it Phys. Rev. Lett.} {\bf 87}
  (2001) 251807, [\href{http://arxiv.org/abs/hep-ex/0108032}{{\tt
  hep-ex/0108032}}].

\bibitem{Limosani:2009qg}
{\bf Belle} Collaboration, A.~Limosani et~al., {\it {Measurement of Inclusive
  Radiative B-meson Decays with a Photon Energy Threshold of 1.7-GeV}},  {\it
  Phys. Rev. Lett.} {\bf 103} (2009) 241801,
  [\href{http://arxiv.org/abs/0907.1384}{{\tt arXiv:0907.1384}}].

\bibitem{Saito:2014das}
{\bf Belle} Collaboration, T.~Saito et~al., {\it {Measurement of the $\bar{B}
  \rightarrow X_s \gamma$ Branching Fraction with a Sum of Exclusive Decays}},
  {\it Phys. Rev.} {\bf D91} (2015), no.~5 052004,
  [\href{http://arxiv.org/abs/1411.7198}{{\tt arXiv:1411.7198}}].

\bibitem{Aubert:2007my}
{\bf BaBar} Collaboration, B.~Aubert et~al., {\it {Measurement of the $B \to
  X_s \gamma$ branching fraction and photon energy spectrum using the recoil
  method}},  {\it Phys. Rev.} {\bf D77} (2008) 051103,
  [\href{http://arxiv.org/abs/0711.4889}{{\tt arXiv:0711.4889}}].

\bibitem{Lees:2012ym}
{\bf BaBar} Collaboration, J.~P. Lees et~al., {\it {Precision Measurement of
  the $B \to X_s \gamma$ Photon Energy Spectrum, Branching Fraction, and Direct
  CP Asymmetry $A_{CP}(B \to X_{s+d}\gamma)$}},  {\it Phys. Rev. Lett.} {\bf
  109} (2012) 191801, [\href{http://arxiv.org/abs/1207.2690}{{\tt
  arXiv:1207.2690}}].

\bibitem{Lees:2012ufa}
{\bf BaBar} Collaboration, J.~P. Lees et~al., {\it {Measurement of B($B\to X_s
  \gamma$), the $B\to X_s \gamma$ photon energy spectrum, and the direct CP
  asymmetry in $B\to X_{s+d} \gamma$ decays}},  {\it Phys. Rev.} {\bf D86}
  (2012) 112008, [\href{http://arxiv.org/abs/1207.5772}{{\tt
  arXiv:1207.5772}}].

\bibitem{Lees:2012wg}
{\bf BaBar} Collaboration, J.~P. Lees et~al., {\it {Exclusive Measurements of
  $b \to s\gamma$ Transition Rate and Photon Energy Spectrum}},  {\it Phys.
  Rev.} {\bf D86} (2012) 052012, [\href{http://arxiv.org/abs/1207.2520}{{\tt
  arXiv:1207.2520}}].

\bibitem{Amhis:2014hma}
{\bf Heavy Flavor Averaging Group (HFAG)} Collaboration, Y.~Amhis et~al., {\it
  {Averages of $b$-hadron, $c$-hadron, and $\tau$-lepton properties as of
  summer 2014}},  \href{http://arxiv.org/abs/1412.7515}{{\tt arXiv:1412.7515}}.

\bibitem{Misiak:2015xwa}
M.~Misiak et~al., {\it {Updated NNLO QCD predictions for the weak radiative
  B-meson decays}},  {\it Phys. Rev. Lett.} {\bf 114} (2015), no.~22 221801,
  [\href{http://arxiv.org/abs/1503.01789}{{\tt arXiv:1503.01789}}].

\bibitem{Li:2013vlx}
X.~Q.~Li, Y.~D.~Yang and X.~B.~Yuan, {\it {Exclusive radiative B-meson decays
 within minimal flavor-violating two-Higgs-doublet models}},
 {\it Phys. Rev.} {\bf D89} (2014) 054024, [\href{http://arxiv.org/abs/1311.2786}{{\tt
  arXiv:1311.2786}}].

\bibitem{Gambino:2003zm}
P.~Gambino, M.~Gorbahn, and U.~Haisch, {\it {Anomalous dimension matrix for
  radiative and rare semileptonic B decays up to three loops}},  {\it Nucl.
  Phys.} {\bf B673} (2003) 238--262,
  [\href{http://arxiv.org/abs/hep-ph/0306079}{{\tt hep-ph/0306079}}].

\bibitem{Lees:2015ymt}
{\bf BaBar} Collaboration, J.~P. Lees et~al., {\it {Measurement of angular
  asymmetries in the decays $B \to K^\ast\ell^+\ell^-$}},  {\it Phys. Rev.}
  {\bf D93} (2016), no.~5 052015, [\href{http://arxiv.org/abs/1508.07960}{{\tt
  arXiv:1508.07960}}].

\bibitem{Capdevila:2017ert}
B.~Capdevila, S.~Descotes-Genon, L.~Hofer, and J.~Matias, {\it {Hadronic
  uncertainties in $B\to K^*\mu^+\mu^-$: a state-of-the-art analysis}},
  \href{http://arxiv.org/abs/1701.08672}{{\tt arXiv:1701.08672}}.

\bibitem{Chobanova:2017ghn}
V.~G. Chobanova, T.~Hurth, F.~Mahmoudi, D.~Martinez~Santos, and S.~Neshatpour,
  {\it {Large hadronic power corrections or new physics in the rare decay
  $B\to K^\ast\mu^+\mu^-$?}},  \href{http://arxiv.org/abs/1702.02234}{{\tt
  arXiv:1702.02234}}.

\end{thebibliography}

\providecommand{\href}[2]{#2}

\end{document}